\documentclass[twocolumn,twocolappendix]{aastex631}
\usepackage{amsmath, mathtools, mathrsfs}
\usepackage{hyperref}
\usepackage{verbatim}
\usepackage{xspace}
\usepackage{longtable}
\usepackage{multirow}
\usepackage{afterpage}
\usepackage{booktabs}
\usepackage{bm}
\usepackage{dsfont}
\usepackage{physics}
\usepackage[thinc]{esdiff}

\defcitealias{sgraI}{\sgra Paper I}
\defcitealias{sgraII}{\sgra Paper II}
\defcitealias{sgraIII}{\sgra Paper III}
\defcitealias{sgraIV}{\sgra Paper IV}
\defcitealias{sgraV}{\sgra Paper V}
\defcitealias{sgraVI}{\sgra Paper VI}
\defcitealias{sgraVII}{\sgra Paper VII}
\defcitealias{sgraVIII}{\sgra Paper VIII}

\def\lsim{\mathrel{\raise.3ex\hbox{$<$\kern-.75em\lower1ex\hbox{$\sim$}}}}
\def\gsim{\mathrel{\raise.3ex\hbox{$>$\kern-.75em\lower1ex\hbox{$\sim$}}}}

\def\m87{M87$^*$\xspace} 
\def\sgra{Sgr~A$^*$\xspace}

\newcommand\edt[1]{{\color{black}#1}}

\begin{document}

\title{Multifrequency Synthesis via CHIBI: Colorful Hierarchical Interferometric Bayesian Imaging}

\author{Erandi Chavez}
\affiliation{Center for Astrophysics $|$ Harvard \& Smithsonian, 60 Garden St, Cambridge, MA 02138, USA}
\author{Paul Tiede}
\affiliation{Black Hole Initiative, Harvard University, 20 Garden St, Cambridge, MA 02138, USA}
\affiliation{Center for Astrophysics $|$ Harvard \& Smithsonian, 60 Garden St, Cambridge, MA 02138, USA}
\author{Sara Issaoun}
\affiliation{Center for Astrophysics $|$ Harvard \& Smithsonian, 60 Garden St, Cambridge, MA 02138, USA}
\affiliation{Black Hole Initiative, Harvard University, 20 Garden St, Cambridge, MA 02138, USA}
\author{Michael D. Johnson}
\affiliation{Center for Astrophysics $|$ Harvard \& Smithsonian, 60 Garden St, Cambridge, MA 02138, USA}
\affiliation{Black Hole Initiative, Harvard University, 20 Garden St, Cambridge, MA 02138, USA}
\author{Dominic Pesce}
\affiliation{Center for Astrophysics $|$ Harvard \& Smithsonian, 60 Garden St, Cambridge, MA 02138, USA}
\affiliation{Black Hole Initiative, Harvard University, 20 Garden St, Cambridge, MA 02138, USA}
\author{Yuh Tsunetoe}
\affiliation{Shanghai Astronomical Observatory, Chinese Academy of Sciences, Shanghai 200030,  P. R. China}
\affiliation{Center for Computational Sciences, University of Tsukuba, \ 1-1-1 Tennodai, Tsukuba, \ Ibaraki 305-8577, Japan}
\author{Daniel C. M. Palumbo}
\affiliation{Black Hole Initiative, Harvard University, 20 Garden St, Cambridge, MA 02138, USA}
\affiliation{Center for Astrophysics $|$ Harvard \& Smithsonian, 60 Garden St, Cambridge, MA 02138, USA}

\begin{abstract}
From magnetized plasma of relativistic jets to dust grains within protoplanetary disks, we study the emission mechanisms of radio sources via their rich spectral structure. Multifrequency Synthesis (MFS) is a technique in which interferometric data at multiple frequencies are imaged simultaneously, resulting in a denser sampling of spatial scales, higher imaging fidelity, and tighter constraints on the source’s spectral structure and evolution. We describe a new method of MFS imaging reconstruction in a hierarchical interferometric Bayesian inference framework, CHIBI. The model parametrization is based on the spectral behavior of synchrotron radiation, the emission mechanism dominating the radio emission observed from galactic nuclei. We show results of this method on observations of jet sources from the MOJAVE catalog with the Very Long Baseline Array, and showcase the prospects for MFS imaging of M87* with simulated data from the Event Horizon Telescope (EHT) and future expansions such as the next generation EHT and the Black Hole Explorer. These demonstrations highlight the benefit of MFS to reconstruct higher-fidelity images and spectral index maps, producing scientifically richer results in a statistically grounded framework, implemented in \texttt{Comrade.jl}.
\end{abstract}


\section{Introduction}

Following the serendipitous discovery of astronomical radio emission by \cite{Jansky1933_Radio}, analysis by \cite{Whipple1937_spectrum} demonstrated that single frequency measurements were insufficient to constrain the physical origin of the observed emission---observations of the source's spectral evolution were needed. This argument is fundamental: astronomical sources are inherently chromatic, and understanding their spectral structure is key to constraining the underlying physical mechanisms powering the observed emission. 

Recent theoretical studies of supermassive black hole accretion flows, especially those with very low accretion rates, suggest an important role for spectral information in measurement of both plasma and spacetime properties. For example, the strongly lensed light in the ``photon ring'' feature of optically thin accretion flow images is notable for its higher synchrotron optical depth, causing a prominent feature in spectral index maps \citep{Ricarte_2023} and an outsized contribution to image-integrated features at high frequencies \citep{Palumbo_2024, Palumbo_2025}. Multifrequency observations are also critical to teasing apart jet-launching structures in the inner accretion flow, where spectral hints at the underlying electron heating mechanisms are most important \citep{YT2025anisotropy}. 

\begin{figure}
    \centering
    \includegraphics[width=1.0 \linewidth]{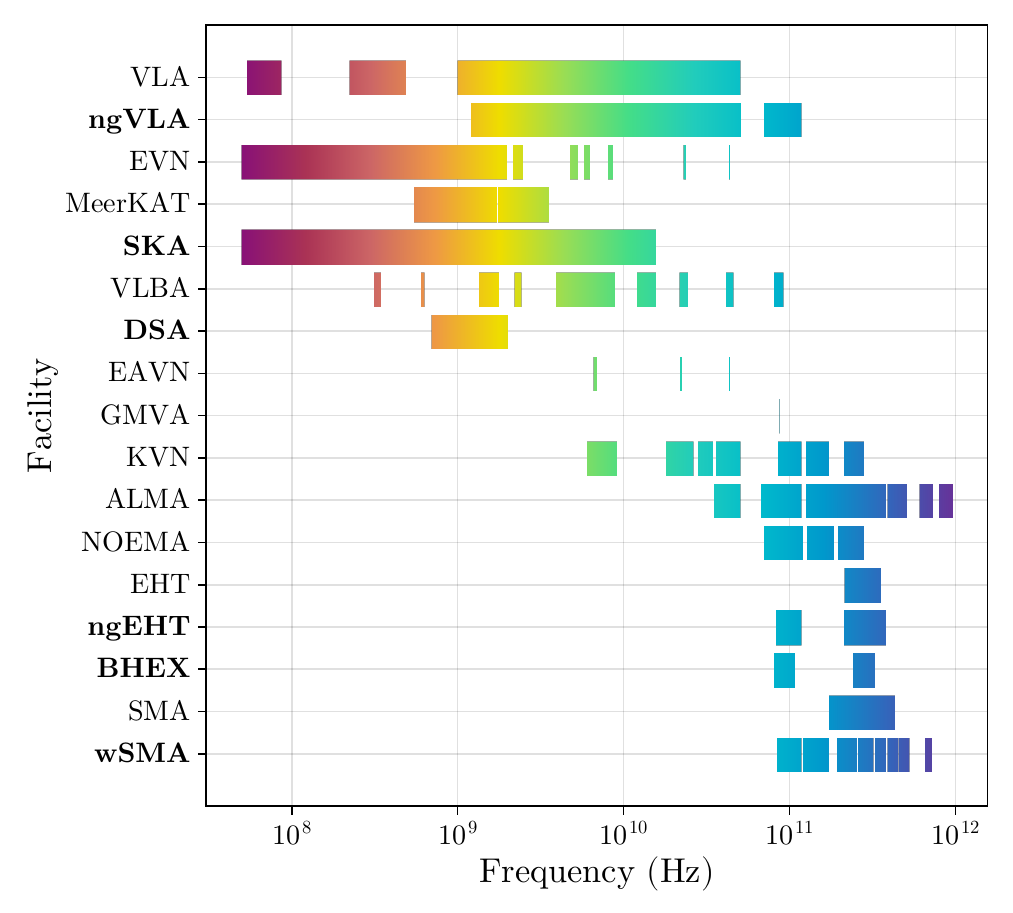}
    \caption{Multifrequency capabilities of various current (unbolded text) and next-generation (bold  text) radio and submillimeter observatories \citep{VLAdocs, ngVLAdocs, M87pII, Doeleman2023_ngEHT, Johnson2024BHEX, VLBAdocs, ALMAdocs, deVilliers2023_MeerKAT, SKAdocs, Hallinan2019_DSA2000, NOEMAdocs, wSMA2024}. Multifrequency imaging is a powerful tool to fully utilize the wide frequency coverage of these facilities.}
    \label{fig:intro}
\end{figure}

Following the decades of technological development since Jansky's original discovery, multifrequency observing has become a fundamental design choice for modern and next-generation instruments, thus shifting the observational landscape to a multi-color view spanning orders of magnitude in frequency (see \autoref{fig:intro}). In pursuit of finer angular resolution at radio wavelengths, many of these facilities are interferometers: radio telescopes that combine simultaneous observations across arrays of radio dishes.

Imaging radio data from very long baseline interferometry (VLBI) is a classic, ill-posed problem. Radio interferometers measure visibilities ($V_{p,q}$), which are the complex cross-correlation of the electric fields ($E$) received at a baseline---formed by two different telescopes $\{p,q\}$ in the array---as a function of time ($t$)\footnote{Note that the times for $E_p$ and $E_q$ are shifted to give equal geometric delay to the source.}, frequency ($\nu$), and polarization ($P$) \citep{TMS}:
\begin{equation}
    V_{p,q}(t, \nu, P) =  \langle E_{p}(t, \nu, P)E_{q}^*(t, \nu, P) \rangle.
\end{equation}
The true, uncorrupted interferometric visibility response, $\mathcal{V}(u,v)$, is related to the on-sky image via a Fourier Transform:
\begin{align}
    \nonumber    \mathcal{V}(u,v; t, \nu, P) &= \iint \, I(x,y; t, \nu, P) \text{e}^{-2 \pi i \left(b_x x + b_y y \right) \nu/c} dx dy\\
    &= \iint \, I(x,y; t, \nu, P) \text{e}^{-2 \pi i (ux+vy)} dx dy.
    \label{eq:idealcomplexvis}
\end{align}
Here, $\{ b_x, b_y \}$ is the physical baseline vector for baseline $\{p,q\}$, projected orthogonal to the line of sight. 

For a given baseline, the relationship between the ideal, uncorrupted visibility ($\mathcal{V}_{p,q}$) and the actual measured visibility ($V_{p,q}$) is:
\begin{equation}
    V_{p,q} \approx g_p\, g_q^{*}\, \mathcal{V}_{p,q} + \epsilon_{pq}. 
    \label{eq:gains}
\end{equation}
Here, $\{g_p, g_q\}$ are complex-valued gain factors at stations $p$ and $q$, respectively, which account for imperfection in the telescope responses and atmospheric corruption of the data. $\epsilon_{pq}$ is the thermal noise.

The goal of VLBI image reconstruction is to invert this Fourier relation to recover $I(x,y; t, \nu, P)$ from limited measurements of $V_{p,q}$.  The sampling of the on-sky image in the visibility domain, or $(u,v)$ coverage, can be increased in multiple ways: making use of Earth-rotation synthesis, adding new stations to the interferometer, or increasing the range of observing frequencies via wider bandwidths and simultaneous receiver systems. 

We are interested in a multifrequency imaging framework that uses observations at a number of frequencies to reconstruct an image of a source along with its spectral characteristics. Our goal is to simultaneously reconstruct multiple frequencies, enforcing physical relationships between them, rather than separately reconstructing images and then comparing them to measure the spectral characteristics. The latter approach is standard and faces a number of challenges, including issues with inhomogeneous baseline coverage and different angular resolution at different frequencies. 

In the simplest scenario, multifrequency parametrization can be done in terms of an image\footnote{From here onward, we proceed with all image reconstructions as static, total intensity images.} $I_{0}$ at some reference frequency $\nu_{0}$ along with a spectral index image $\alpha$. The image at any other frequency is given by:
\begin{align}
    I(x, y; \nu) = I_{\rm 0}(x,y; \nu_{0}) {\left( \frac{\nu}{\nu_{0}} \right)^{\alpha(x, y)}}.
\end{align}
This simple parameterization is physically motivated: blackbody radiation in the Rayleigh-Jeans limit obeys a power law, and synchrotron radiation and bremsstrahlung both obey power laws in the optically thin and thick limits. 
With this prescription, the imaging problem becomes simultaneously reconstructing \emph{two} images, $I_{0}(x,y; \nu_{0})$ and $\alpha(x, y)$, subject to multifrequency data constraints. 

CLEAN \citep{Hogbom1974CLEAN} is the standard imaging method for interferometric data. Traditional CLEAN is a deconvolutional method, in which the interferometer's point spread function is deconvolved from the image by iteratively modeling the on-sky image as a collection of point sources. The final step is blurring the resulting image with a Gaussian kernel the size of the interferometer's resolution. There exist a number of CLEAN-based algorithms developed to simultaneously fit data at multiple observing frequencies and produce images and spectral index maps \citep{Sault_1994,Rau_2011,Offringa_2017}. 

In contrast to CLEAN, an alternative approach to VLBI imaging is to forward model the visibility response given a pre-defined image model---typically a rasterized image \citep{Frieden1972_MEM,NarayanandNityananda1986_MEM, Wiaux2009_MEMI,Wiaux2009_MEMII}. Regularized maximum likelihood (RML) is one such method, which forward-models the observed data from an underlying image model by maximizing an objective function weighted by data and image prior constraints \citep[e.g.,][]{Akiyama_2016,Chael18_Closure,M87PaperIV}. Recently, \citet{Chael_2023} developed a multifrequency extension of the RML imaging approach, simultaneously fitting a reference frequency image, spectral index map, and potentially a spectral curvature map. The lack of a blurring step during imaging results in a ``super resolution'' effect, in which structure on scales smaller than the instrument resolution is recovered.
Such an imaging approach however relies on finding a single maximum of the RML function. A quantitative assessment of image fidelity with this method requires extensive and computationally expensive parameter surveys \citep[e.g.,][]{M87PaperIV,SgrAPaperIII}. 

Other algorithms have been developed to perform full Bayesian modeling of the image pixels, and therefore estimate full posterior probability distributions in the image fitting \citep{Sutter2014_gibbs, Junklewitz2016_resolve,Broderick2020_THEMIS,hibi}. 
\edt{Applications of Bayesian modeling with multifrequency fitting have already been published, such as the imaging of spectral components of the \textit{Fermi} LAT gamma-ray sky \citep{ScheelPlatz_2023}, imaging of modest spectral variations in the M87* ring structure from Event Horizon Telescope (EHT) 2017 observations \citep{Arras2022}, and imaging of simulated observations expected from EHT instrument and telescope upgrades in the next decade \citep{Roelofs_2023}.}  
These Bayesian inference techniques also benefit from super-resolution and sample the full image posterior. The sampled posterior provides quantified uncertainties on the resulting image features, which is critical to evaluate imaging fidelity of reconstructions produced from sparse VLBI data. 

In this paper, we extend a hierarchical interferometric Bayesian imaging approach \citep[HIBI;][]{hibi} to multifrequency synthesis, where the image prior is based on Markov random fields. \edt{By utilizing Markov random fields, our approach to Bayesian imaging is simpler than previous methods, e.g., \citet{Arras2022, Roelofs_2023}. Due to their simple hierarchical structure, HIBI and its multifrequency extension use Monte Carlo methods for inference rather than Gaussian approximations, which can be inaccurate for sparse arrays such as the EHT or VLBA.} Our associated code is implemented in the \texttt{Comrade.jl} software repository for HIBI \citep{tiede2022}.

\section{Method Description}

\begin{figure*}
    \centering
    \includegraphics[width=1.0\linewidth]{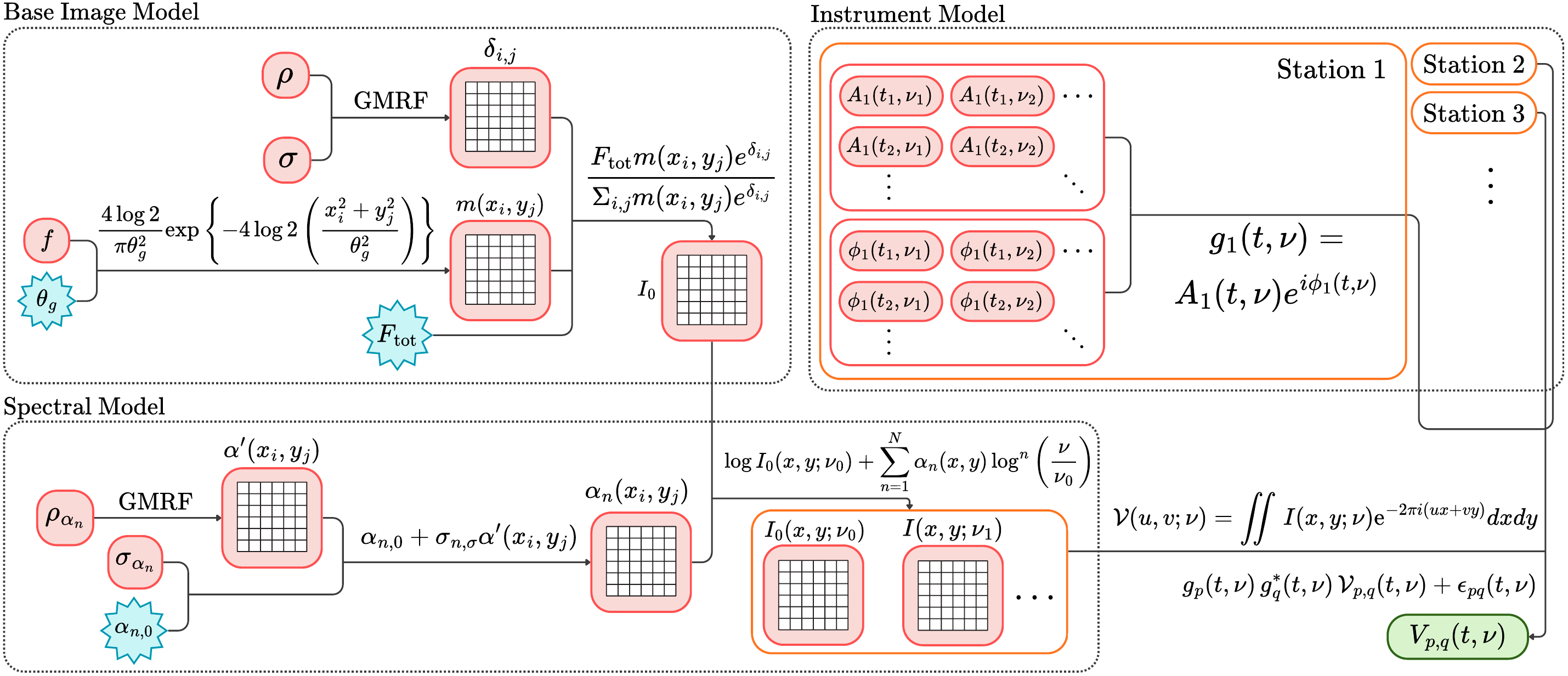}
    \caption{\edt{Schematic outlining the full hierarchical forward model as implemented in CHIBI when fitting complex visibilities as the data product. Parameters in blue stars represent those fixed by the user in advance. The grid parameters (FOV$_x$, FOV$_y$, N$_x$, N$_y$), not shown here, are also fixed by the user.}}
    \label{fig:CHIBI_model}
\end{figure*}

\subsection{Multifrequency Model\label{sec:model}}
Multifrequency synthesis requires a spectral distribution model that relates the image structure across different frequencies. We chose a model parametrization grounded in the emission mechanism powering our sources of interest: synchrotron radiation. 

In the radio, synchrotron emitters commonly appear as power laws, where the slope of this power law in $\log(I)-\log(\nu)$ space is the spectral index. Explicitly modeling the spectral index is critical as it encodes the underlying electron distribution function of the emitting plasma. Modeling higher-order structure is also of interest because changes in the spectral slope can indicate either optical depth effects, or multiple electron populations in the source (e.g., multiple emitting regions). This emission mechanism connects to past research \citep{Conway1990, Junklewitz2015_resolveMFS, chael2023ehtim} and motivates our choice of spectral model: a polynomial expansion in $\log{I} - \log{\nu}$ space about a reference image $I_0$ at a frequency $\nu_0$. We adopt the spectral model prescription from \cite{Chael_2023}:
\begin{equation}\label{eq:polyspectral}
    \begin{aligned}
        \log{I}(x,y; \nu) = \: &\log I_0(x,y; \nu_0)\\
        &+ \sum_{n=1}^N \alpha_n(x,y) \log^n{\left(\frac{\nu}{\nu_0}\right)}.
    \end{aligned}
\end{equation}
We have implemented this model such that the expansion can be performed to an arbitrary order. The first and second-order coefficients, $\alpha\equiv \alpha_1$ and $\beta\equiv \alpha_2$, correspond to the spectral index\footnote{We define the sign of spectral index as $\alpha \equiv +\diff{\log{I(\nu)}}{\log{\nu}}$} and spectral curvature, respectively  \citep[e.g.][]{Kellermann1964_spectralindex,Baars1977_curvature}. Since different regions of the image can have different emitting populations, we make the coefficients of the expansions, e.g., the spectral index and curvature, a function of the image coordinates.

For computational purposes, we rasterize our reference image $I_0$ and the spectral coefficients ($\alpha$, $\beta$, ...) on a grid of pixels (each of which has a tunable flux) of fixed size and dimension $N_x\times N_y$ with field of view $({\rm FOV_x, FOV_y})$. For the reference image, we use the decomposition
\begin{equation}
    I_0(x_i, y_j) = F_{\rm tot} \frac{m(x_i, y_j)e^{\delta_{i,j}}}{\sum_{i,j} m(x_i, y_j)e^{\delta_{i,j}}},
    \label{eq:totalflux}
\end{equation}
where $m$ encodes the a priori information (see \autoref{sec:priors} for our choice of image prior) about the image and $e^{\delta_{i,j}}$ are the multiplicative fluctuations about the structure $m$. $F_{\rm tot}$ parameterizes the total flux of the reference image, which can either be fixed a priori (e.g. in the case of existing total flux measurements) or fit as an additional model parameter. All image tests here have image models with fixed total flux. We parameterize $m(x,y)$ as a constant floor with an added Gaussian component of FWHM $\theta_g$:
\begin{equation}
    m(x,y) = (1-f) m_g(x,y; {\rm \theta_g}) + \frac{f}{N_xN_y},
\end{equation}
where $f$ models the fraction of emission placed in a constant background flux, and $m_g$ is given by
\begin{equation}\label{eq:mug}
    m_g(x,y; \theta_g) = \frac{4 \log(2)}{\pi \theta_g^2}e^{-4\log(2) (x^2 + y^2)/\theta_g^2}.
\end{equation}

Our choice for the value of $\theta_g$ depends on whether or not we fit for instrumental gains. If we solely solve for the image structure (no instrument modeling), we use $\theta_g = \max \{{\rm FOV}_x, {\rm FOV}_y\}/2$. This places a slight preference for concentrating flux towards the center of the image grid, but ultimately is not very informative regarding specific source structure. When we do perform instrument modeling and fit for instrumental gains, the image structure can drift during the posterior sampling process (see \autoref{sec:degeneracies} for more details on the underlying degeneracy). In this case, we choose a smaller $\theta_g$---typically $\theta_g = \max \{{\rm FOV}_x, {\rm FOV}_y\}/10$ or smaller---to concentrate the flux in one location of the image. Although not completely eliminated, a more restrictive mean image structure greatly reduces drifting effects during the sampling process. 

We model the corresponding spectral maps, $\alpha_n(x,y)$, as raster images with identical dimensions as the corresponding reference image. For the spectral maps we do not use the same parametrization as the reference image (\autoref{eq:totalflux}), instead our model parameters are the pixel values of the raster image:
\begin{equation}
    \alpha_n(x_i,y_j).
\end{equation}
We also show an alternative parameterization in which we explicitly parameterize an offset in the spectral map (see \autoref{fig:ngEHTsummary} for motivation and results):
\begin{equation}
    \alpha_n(x_i,y_j) = \alpha_{n,0} + \sigma_{n,\alpha} \alpha_n'(x_i,y_j),
\end{equation}
where $\alpha_n'(x_i,y_j)$ are the gridded pixel values.

We are capable of modeling the station-dependent gain factors in \autoref{eq:gains} alongside the image structure. For station $p$, we model the gain factor at frequency $\nu$ as
\begin{equation}
    g_p(t; \nu) = e^{\ln A_p(t; \nu) + i\phi_p(t; \nu)},
    \label{eq:gainmodel}
\end{equation}
where $\ln A_p$ is the natural log of the gain amplitude, and $\phi_p$ is the gain phase. Fitting the natural log of the gain amplitude ensures the gaim amplitude itself always remains positive. We do not introduce any frequency-correlation between the gains, we assume that the gain behavior across frequencies is independent. We demonstrate tests with and without instrument modeling within this paper.

\subsection{Model Priors\label{sec:priors}}

For the image fluctuations $\delta_{i,j}$ and spectral map raster $\alpha_n(x_i, y_j)$ we use a zero-mean Gaussian Markov random field (GMRF) prior
\begin{equation}\label{eq:GMRF}
\begin{aligned}
                p(\bm{r} | \bm{Q}) &= \sqrt{\frac{\det \bm{Q}}{2\pi^K}}
            \exp\left(-\frac{1}{2}\mathbf{r}^T \bm{Q}\, \bm{r}\right),
\end{aligned}
\end{equation}
where $\bm{r}$ is the random field realization and $\bm{Q}$ is a sparse precision matrix (the inverse covariance matrix). We will denote samples from this distribution as
\begin{equation}
    \bm{\delta} \sim {\rm GMRF}(\bm{0}, \bm{Q}),
\end{equation}
For our precision matrix, we follow \citet{hibi} and use a first-order random field 
\begin{equation}\label{eq:first-order-GMRF}
    \bm{Q}_1 = \frac{1}{\sigma^2}\left( \frac{1}{\rho^2}\mathds{1} + \bm{G}\right),
\end{equation}
where $\mathds{1}$ denotes the identity matrix and $\bm{G}$ denotes
\begin{equation}\label{eq:0orderGMRF}
    G_{ij, kl} = \begin{cases}
        4  &  ij=kl\\
        -1 & ij \sim kl\\\ 
        0  & {\rm otherwise}
    \end{cases},
\end{equation}
where $ij\sim kl$ means that on the image pixel grid $ij$ is adjacent to $kl$. \citet{Besag1981_system} demonstrated that this is equivalent in the large raster limit to assuming that the stochastic field has the spectral density 
\begin{equation}
    S(\bm{k})\dd^2\bm{k} \propto \frac{\dd^2\bm{k}}{1 + \rho^2\norm{\bm{k}}^2},
\end{equation}
where $\bm{k}$ is the spatial frequency.

For large spatial scales ($\norm{\bm{k}} \ll 1/\rho$), the power spectrum is constant and behaves like a white noise process. However, the power spectrum contains a break at $\norm{\bm{k}} \sim 1/\rho$. For spatial scales smaller than $\rho$, ($\norm{\bm{k}} \gg 1/\rho$), the power spectrum transitions into a power law:
\begin{equation}
    S(\bm{k})\dd^2\bm{k} \propto \norm{\bm{k}}^{-2}\dd^2\bm{k},
\end{equation} 
which suppresses spatial structures below $\rho$. This suppression motivates the interpretation of $\rho$ as the GMRF correlation length, which we model as a hyperparameter. 

We use an inverse gamma distribution ($\mathcal{IG}$) as the prior for the correlation length:
\begin{equation}\label{eq:corr_prior}
    \rho \sim \mathcal{IG}\left[1, -\log(0.01)\lambda_{\rm  beam}\right],
\end{equation}
where $\lambda_{\rm beam} = 1/{\rm max}(\norm{\bm{u}})$ is roughly the beam size of the telescope. This assigns only 1\% of the prior mass on $\rho$ to a correlation length below the telescope's beam size. Super-resolved structure in our images is not preferred by the prior, therefore must be driven by the likelihood---the data. For a deeper discussion of the GMRF prior and alternative image priors for Bayesian hierarchical imaging, see \citet{hibi}. 

The global standard deviation $\sigma$ for the GMRF is also fit for as a hyperparameter. We use a half Normal distribution ($\mathcal{HN}(0,0.2)$) with mean 0, standard deviation 0.2, and a lower bound of 0 for the prior on this $\sigma$ parameter. 

This GMRF prior is used for the image fluctuations applied to the reference image $\delta_{i,j}$, and the spectral map raster grids, $\alpha_n(x_i,y_j)$:
\begin{equation}
\begin{aligned}
    \bm{\delta} &\sim {\rm GMRF}(0, \bm{Q}_1),\\
    \bm{\alpha_n} &\sim {\rm GMRF}[0, \bm{Q}_1(\sigma=1)].
\end{aligned}
\end{equation} For the total intensity and each spectral map, we fit a unique value of $\rho$ for each map---no assumptions are made to correlate the value $\rho$ between $\delta(x,y)$ and $\alpha_n(x,y)$. However, for the spectral maps' priors we no longer fit for the global standard deviation $\sigma$ and instead assume a value of $\sigma = 1$. For the tests in this paper where only fit up to first order spectral structure ($\alpha(x,y)$), this places a preference of order $\sim 1$ values for spectral index. This assumption is motivated by observational evidence of non-thermal radio sources \citep[e.g.][]{Kellermann1964_spectralindex} and theoretical predictions of relativistic AGN jets \citep{BlandfordKonigl1979_core_jet}, accreting black hole systems \citep{Ricarte_2023}, and shocks \citep{Bell1978_shocks,BlandfordandOstriker1978_shocks} in the radio.

For all tests shown here, the prior for station $p$'s log-gain amplitude ($\ln A_p$) is a Gaussian with mean 0 and standard deviation 0.2:
\begin{equation}
    \bm{{\rm ln}A_p(t; \nu)} \sim \mathcal{N}(0, 0.2).
\end{equation}
We parameterize the station phase priors as a von Mises ($\mathcal{VM}$) distribution with a mean of 0 and a concentration parameter $\kappa = 1/\pi^2 \sim 0.1$:
\begin{equation}
    \bm{\phi_p(t; \nu)} \sim \mathcal{VM}(0, 1/\pi^2).
\end{equation}
This functions as a nearly-uniform, wrapped prior on the phases with a very slight preference for $\phi_p$ around 0. We make no assumptions about the relations between gain parameters (amplitude or phase) between different stations, or between different frequencies. This could be improved for arrays which observe multiple frequencies simultaneously, which are likely to have correlated, frequency-dependent gain behavior due to correlated atmospheric corruption across frequencies \citep[which can be leveraged with instrumental techniques such as Frequency Phase Transfer, e.g.][]{Rioja_2020}, and similar instrument pathways. \edt{The full hierarchical forward model implemented in CHIBI is shown in \autoref{fig:CHIBI_model}.}

\subsection{Initialization and Data Products \label{sec:initialization_data_products}}

All image tests shown here are initialized via a random draw from the priors. After initialization, we use a stochastic gradient descent algorithm \citep[Adam;][]{adam} to traverse the parameter space to where the bulk of the posterior is located. This is necessary because random draws from the prior typically do not lie in high probability regions of parameter space. After reaching a region in the posterior where $\chi^2 \sim 1$, we stop the optimization algorithm and switch to HMC---specifically the NUTS algorithm \citep{NUTS}---which samples the posterior. Each sample from the posterior corresponds to a single image solution---sampling the posterior results in many possible image solutions to our data. Therefore, unless otherwise noted, all of our imaging results here are mean images from the posterior, where this mean is calculated by a pixel-by-pixel averaging of our samples in image-space. 

In the case of a multi-modal posterior (multiple, distinct image structures consistent with the data e.g. two flipped versions of the same image structure, see \autoref{sec:degeneracies} and \autoref{sec:solutions_multimodal} for more discussion) NUTS may fail to find these multiple modes during a single sampling run. Multiple, distinct initializations are typically necessary to identify these different modes.

For all image tests in this paper, we fit complex visibilities as our data products. 
Other than for a limited number of validation runs, we fit for station gains (e.g., instrumental and atmospheric corruptions) during all of the synthetic- and real-data imaging tests presented in this paper. Note that when we fit for instrumental gains, we use the more restrictive value of $\theta_g$, the mean image structure, described in \autoref{sec:model} and \autoref{sec:degeneracies} (gaussian with a smaller FWHM) to minimize the impact of image drift.

\subsection{Method Validation} \label{sec:validation}
To validate this multifrequency model, we constructed a synthetic data test comprised of a single Gaussian at two observing frequencies: 8.1 and 12.1 GHz. These are the two observing frequencies from the VLBA MOJAVE OJ287 dataset in this paper (see \autoref{sec:oj287}) which we used to generate the $(u,v)$ sampling and noise parameters for this test. The image structure at each frequency is a circular gaussian of $\text{FWHM} = 1.5\,{\rm mas}$ centered at the image origin. The total fluxes ($F_\text{tot}$) of the Gaussian sources at each frequency are:
\begin{align*}
    \textrm{8.1 GHz: } F_\text{tot} &= 1.2 \textrm{ Jy},\\
    \textrm{12.1 GHz: }F_\text{tot} &= 1.6 \textrm{ Jy}.
\end{align*}
We chose a Gaussian source as our image structure for ease of interpretation in the image and visibility domains. The spectral evolution of this source corresponds to a constant spectral spectral index of $\alpha(x,y) \equiv \log(1.6/1.2)/\log(12.1/8.1) \simeq 0.716.$ with no higher order spectral structure $\beta(x,y) \equiv 0$.

For this dataset, we performed multifrequency and single frequency image tests for a direct comparison of the code performance with the established HIBI method. All these validation tests fit complex visibilities as the data products. For the multifrequency and single frequency fits, we performed two imaging tests: one where we solely model the image structure, and one where we model image structure \textit{and} the station-based gains. This is critical because real data typically contains residual calibration errors which can corrupt the image structure. Joint image and instrument modeling allows for the model to correct for these residual errors and fold the corresponding calibration uncertainty into the sampled posterior. Therefore we validate the method's performance with imaging alone and in the more appropriate regime for real data: image modeling in conjunction with instrument modeling. 

Here, we describe the imaging model we used for these various imaging tests. All four validation tests used the standard imaging model described in \autoref{sec:model}. All tests had an image field of view (FOV) of $10 \times 10\,\text{mas}$ and image dimensions of $64 \times 64 \,\text{pixels}$ $(\sim 3.8 {\rm \,pixels/beam})$. For the mean image structure parameter $\theta_g$, we use $\theta_g = 5\,\text{mas}$. For the multifrequency tests, we fit up to first order spectral structure $\{I_0(x,y)$, $\alpha(x,y)\}$ with 8.1 GHz as the reference frequency. For the tests where we performed joint image and instrument modeling, we use the instrument model described in \autoref{sec:model} with the standard prior choices described in \autoref{sec:priors}. 

\begin{figure*}
    \centering
    \includegraphics[width=0.85\linewidth]{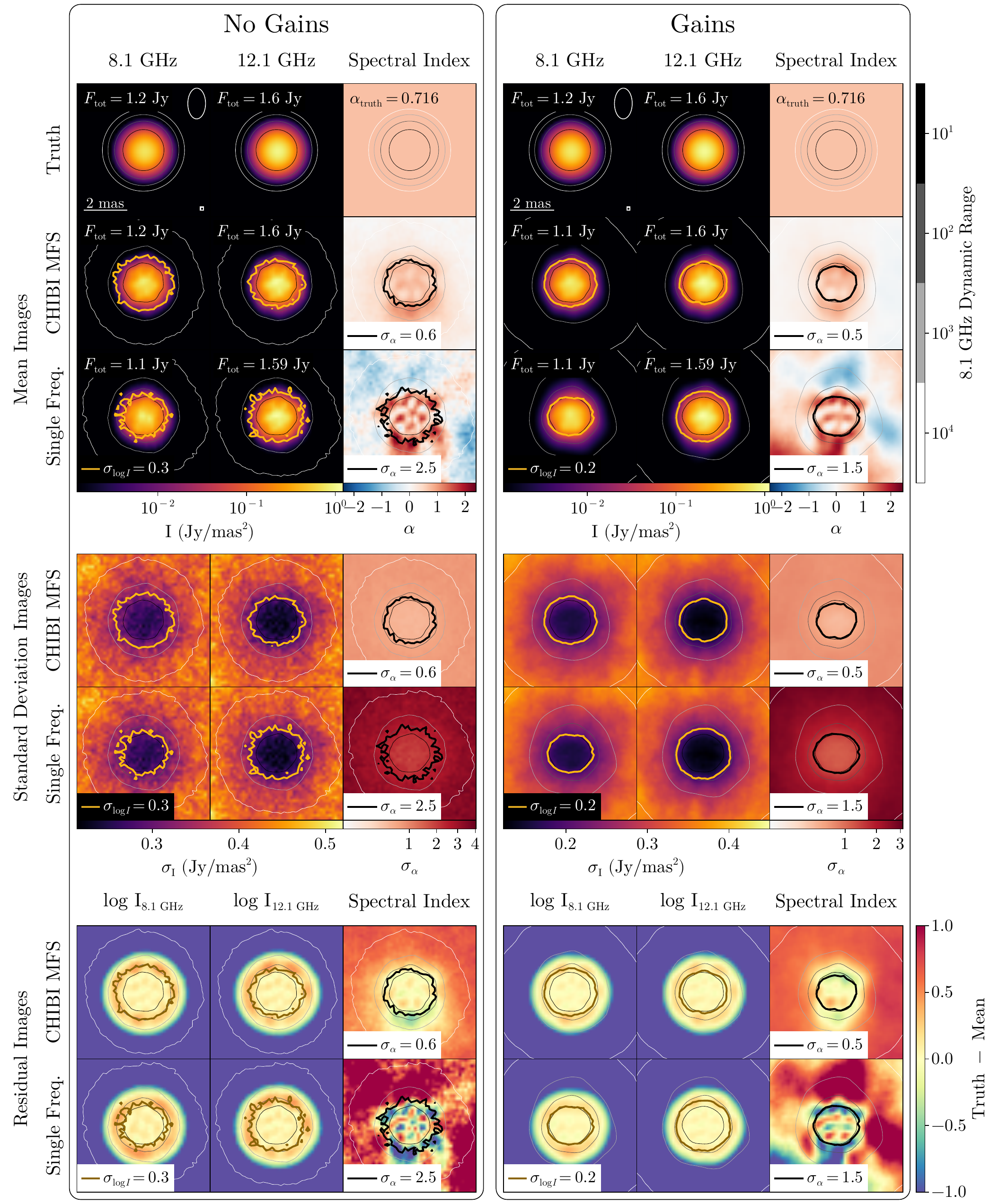}
    \caption{\edt{A comparison between single frequency (HIBI) and multifrequency CHIBI imaging of a multifrequency Gaussian source. The $(u,v)$ coverage for this test is the VLBA coverage for the MOJAVE OJ287 test (see \autoref{fig:uvcoverage}). The topmost row shows the ground truth images. Each subsequent grouping of rows shows the imaging results for each plotted quantity. The top grouping shows the posterior mean total intensity and spectral index images. The middle grouping shows the standard deviation of the log-scaled total intensity $\sigma_{\log I}$ and the spectral index $\sigma_{\alpha}$. The bottom grouping shows the residual images: Truth - Mean. From left to right, the columns represent the 8\,GHz image, 12\,GHz image, and spectral index results. The thin grayscale contours are identical across rows---they show the corresponding logarithmically spaced 8\,GHz intensity contours. The bold lines on the total intensity (gold) and the spectral index (black) results are ``credibility contours'': all regions within this contour have an uncertainty less than the indicated value. We overplot the 8.1\,GHz beam and the reconstruction pixel size on the top left plot for comparison---note that no images shown here are blurred. As expected, the credibility contours are asymmetric because of asymmetric baseline coverage, with the major axis oriented roughly along the minor axis of the beam.}}
    \label{fig:gaussmfvis}
\end{figure*}

\begin{figure*}
    \centering
    \includegraphics[width=1.0\linewidth]{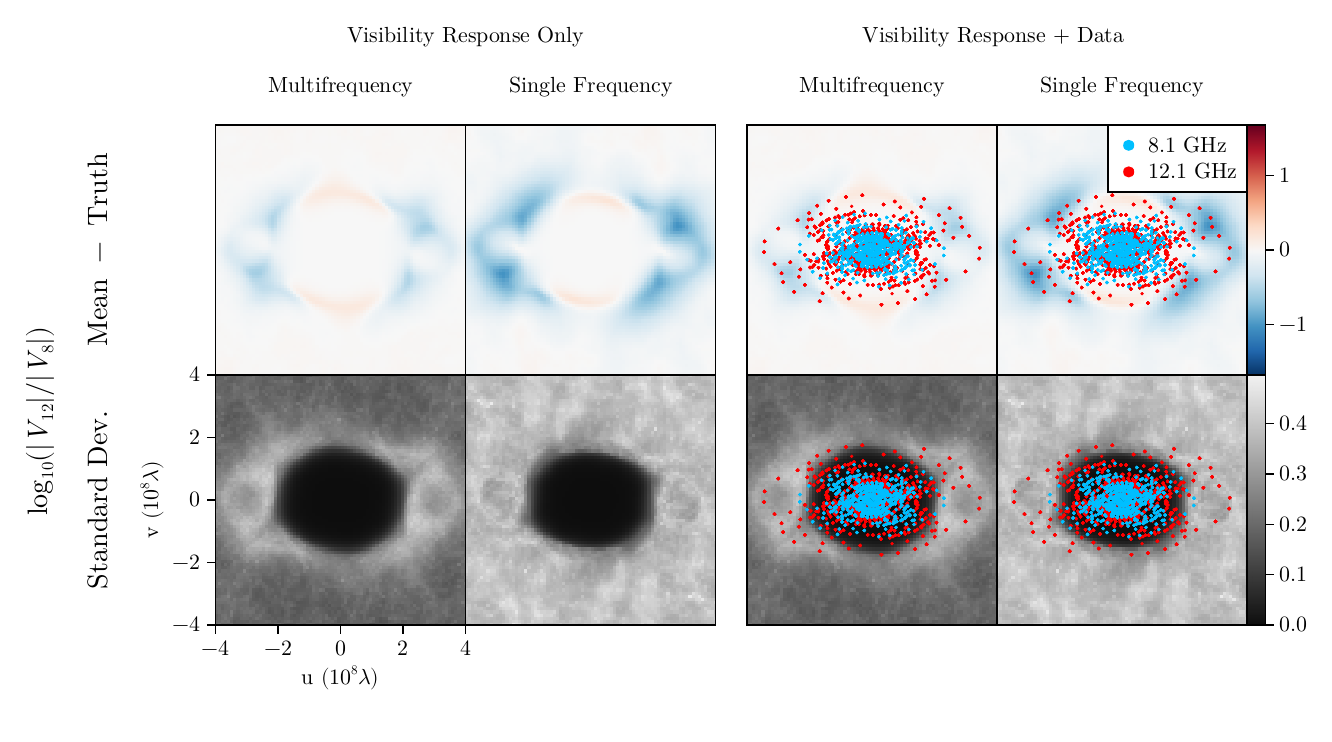}
    \caption{Results from the Gaussian synthetic data tests (without modeling station-based gains) in the visibility domain (see \autoref{fig:gaussmfvis} for the image-domain results). Here, we show the ratio between the visibility response at 12 GHz and 8 GHz in log-space: $\log_{10}(|V_{12}|/|V_{8}|)$. This quantity traces the spectral evolution of this synthetic data test in visibility space---in the test shown this is purely driven by the spectral index because we do not fit for higher-order spectral structure. The top row shows the residuals between ground truth and the mean of $\log_{10}(|V_{12}|/|V_{8}|)$ across the posterior. The left and right plot blocks are identical, the right plot block has the $(u,v)$ coverage over-plotted to guide the eye. For each plot block, the left and right columns correspond to the multifrequency (CHIBI) and single frequency imaging results, respectively. CHIBI recovers the multifrequency structure at all scales more accurately. The regions of visibility space which show the most improvement in accuracy (e.g. smaller residuals) when moving to single frequency to MFS imaging are the regions which have multifrequency data coverage. The existence of an explicit spectral prior from the multifrequency fit results in a modest suppression of uncertainty in high-frequency structure beyond the VLBI sampling. However, between both methods the regions in visibility space with lowest residuals are where \textit{both frequencies} of data are present. MFS is transferring information between frequencies and produces more accurate results than single frequency imaging, however, it is not a substitute for overlapping $(u,v)$ multifrequency coverage.}
    \label{fig:gaussvis}
\end{figure*}

The results from the \edt{validation test with and without joint instrument modeling are shown in \autoref{fig:gaussmfvis}}. We also show the results from the image-only tests in the visibility domain in \autoref{fig:gaussvis}.

Due to the loss of absolute position information in the imaging problem when performing instrument modeling (see \autoref{sec:discussion} for more details), the image structure can drift during posterior sampling. To remove this drifting effect in single frequency imaging, we center each image of the posterior with respect to their centroids. In the case of multifrequency imaging, we center both the reference image and the corresponding spectral index map with respect to the \textit{reference image's} centroid.

We also include ``credibility contours'' for each image test in this paper. These contours represent a credible region within the image---structure within this contour is better constrained than the structure outside. Here, we will use $\left\{ I^{(s)} \right\}_{s=1}^{S}$ to denote the collection of $S$ samples of $I$ from the posterior. For total intensity images, the credibility contours are calculated by taking the pixel-wise standard deviation ($\text{std}$) of the \textit{logarithmically-scaled} total intensity images:
\begin{equation}
    \sigma_{\log_{10} I}(x,y)
= \operatorname{std}\!\left(\left\{\log I^{(s)}(x,y)\right\}_{s=1}^{S}\right).
\end{equation}
Calculating the standard deviation in log-space is a natural choice because the image fluctuations $\delta^{i,j}$ are also sampled in log-space (see \autoref{eq:totalflux}), and these fluctuations encode any image structure deviating from the mean image, $m(x,y)$.
For the spectral index maps, the credible contours are calculated from the pixel-wise standard deviation of the spectral index posterior samples:
\begin{equation}
    \sigma_{\alpha}(x,y)
= \operatorname{std}\!\left(\left\{ \alpha^{(s)}(x,y)\right\}_{s=1}^{S}\right).
\end{equation}

For both our imaging scenarios (with or without modeling station-based gain factors), multifrequency imaging is able to recover the ground truth total intensity images and  spectral index maps. In both scenarios, the multifrequency imaging results are superior to the single-frequency imaging. The improved $(u,v)$ coverage from combining multiple frequencies of data and the spectral index map prior improve the accuracy and precision of the multifrequency imaging results over the single frequency approach. The primary difference between the imaging test without gains and with gains is that the imaging results appear to be smoothed in comparison to the no-gains test. This is due to the image centering procedure described earlier; slight differences between posterior samples due to imperfect image centering functions as a spatial-averaging over the image structure, which smoothens the image results.

\section{Applications}\label{sec:validation}

\begin{figure}
    \centering
    \includegraphics[width=1.0\linewidth]{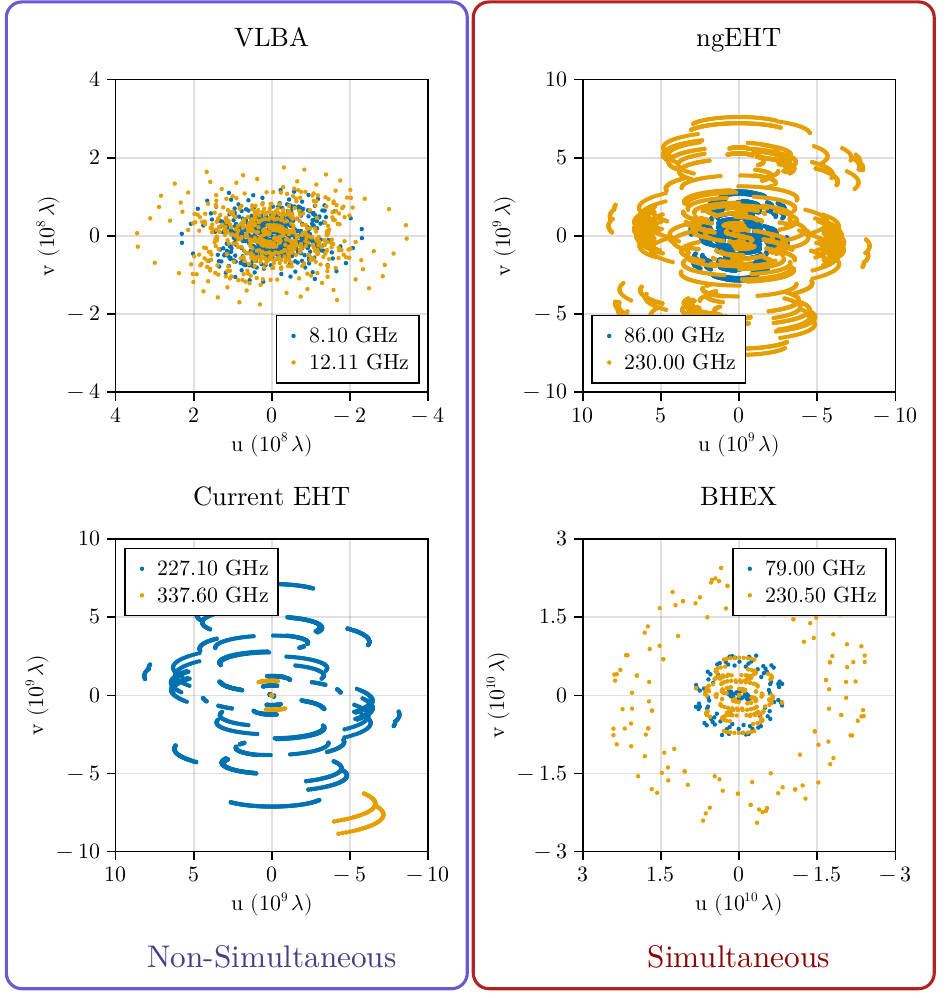}
    \caption{ Multifrequency ($u$,$v$) plots for each of the four arrays featured in this paper. these represent four different frequency-pair scenarios for multifrequency synthesis. The VLBA offers non-simultaneous multifrequency synthesis with very similar coverage. The ngEHT offers dense, multi-scale, and simultaneous frequency coverage. The current EHT multifrequency coverage is sparse, such that the coverage at the higher frequency is not imageable without multifrequency synthesis. BHEX offers simultaneous multifrequency coverage, where the space-ground baselines are significantly longer than the ground array. All of these properties lead to different challenges in multifrequency synthesis. }
    \label{fig:uvcoverage}
\end{figure}

After validation, we applied CHIBI to four different imaging tests that represent different regimes of $(u,v)$ coverage and VLBI observing modes. We perform a real data test on MOJAVE multifrequency observations of the blazars OJ287 (at 8.1 and 12.1 GHz) and PKS 1424+240 (at 12, 23, and 43 GHz), then we perform a series of synthetic data tests of a GRMHD simulation of M87 with three different next-generation VLBI arrays. These three synthetic tests are with the current EHT, the next-generation EHT (ngEHT), and with a ground array combined with the Black Hole Explorer (BHEX) space telescope. Of these four tests, the real VLBA data and the current EHT tests represent a quasi-simultaneous multifrequency observing mode, while the ngEHT and BHEX data are generated with simultaneous multifrequency observing capabilities (see \autoref{fig:uvcoverage}).

The VLBA $(u,v)$ coverage represents a dense overlap between multiple frequencies. The EHT multifrequency coverage is extremely sparse, and represents the scenario where imaging at one frequency is impossible with single frequency imaging (see \autoref{sec:EHT}). The ngEHT coverage represents dense, continuous multifrequency coverage where one frequency's coverage extends past the other frequency's coverage significantly. BHEX represents the case of space VLBI, where the annuli of $(u,v)$ coverage contributed by the ground-space baselines cause very little to no overlap between the two frequencies' sampling.

\subsection{MOJAVE: OJ287} \label{sec:oj287}

The BL Lacertae object OJ287 is known for its 12-year quasi-periodic optical outbursts \citep[e.g.,][]{Sillanpaa_1988,Villata_1998} and jet precession \citep{Britzen_2018}, commonly interpreted as evidence for a supermassive black hole binary \citep{Lehto_1996, Valtaoja_2000, Valtonen_2008}. This source has been regularly monitored with the VLBA for over two decades a part of the MOJAVE monitoring program \citep{Lister2018MOJAVE}. We chose OJ287 because its structure within the published MOJAVE results fits the classic core-jet theoretical radio emission prediction from \cite{BlandfordKonigl1979_core_jet}: an unresolved optically thick core with positive spectral index, followed downstream by optically thin and diffuse jet emission with negative spectral index. 

The OJ287 observations shown were taken with the VLBA at 8.1 and 12.1 GHz on April 28, 2006 as part of the MOJAVE monitoring program \citep{Lister2018MOJAVE}. Note that 8.4 and 15 GHz data from this epoch are also available, but we did not analyze them here. CLEAN spectral index maps and total intensity images of these data sets were published in \citet{Hovatta2014_MOJAVE_spectral_index} and \citet{Lister2018MOJAVE}, respectively. 

\begin{figure*}
    \centering
    \includegraphics[width=1.0\linewidth]{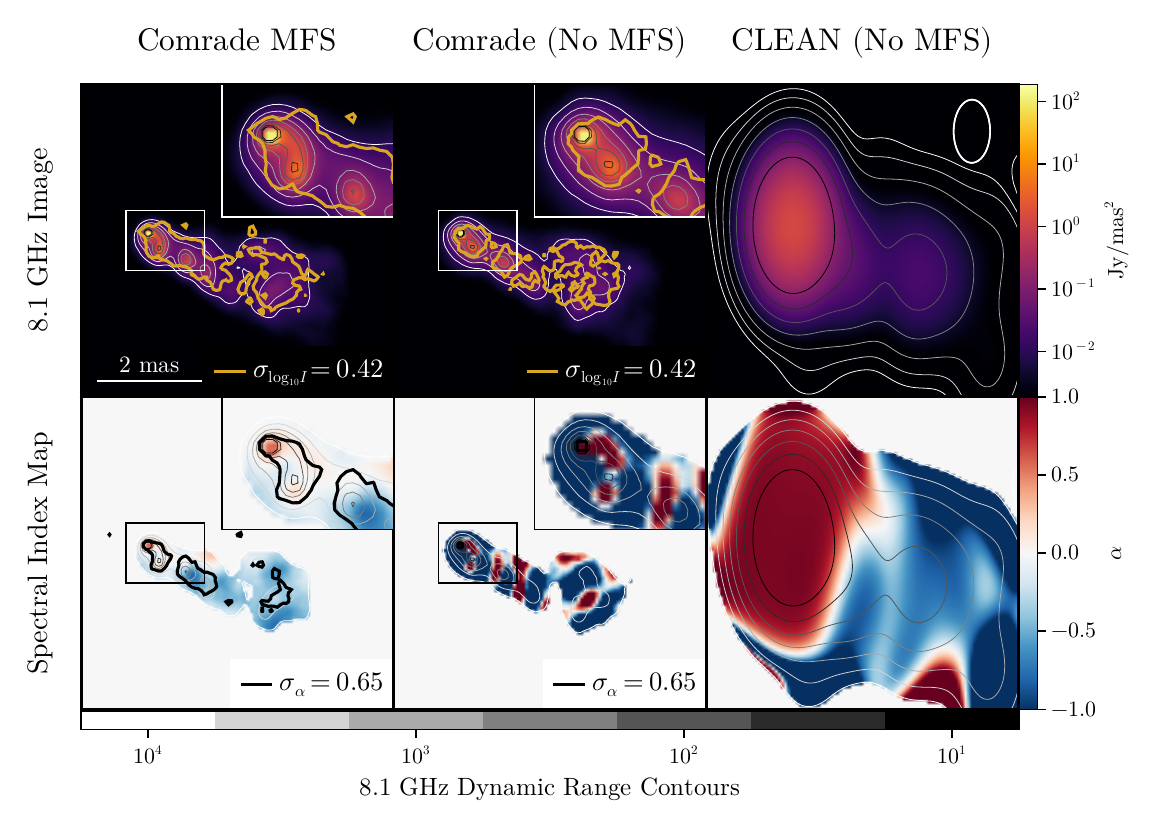}
    \caption{OJ287 reconstructions from VLBA MOJAVE data at 8.1 and 12.1\,GHz. From left to right, we show the multifrequency imaging results (CHIBI), the single frequency imaging results (HIBI), and the published single frequency CLEAN imaging results of this data \citep{Hovatta2014_MOJAVE_spectral_index,Lister2018MOJAVE}. The top row shows total intensity image at the reference frequency (8.1 GHz) and the bottom row shows the corresponding spectral index map for each test. The thin, grayscale contours are identical for each column; they trace out total intensity emission of the reference image in log-scale. The bold lines on the \texttt{Comrade.jl} images represent the ``credibility contours'' calculated from the posterior for this test: the gold lines trace $\sigma_{\log I} = 0.42$, and the black lines trace out $\sigma_{\alpha} = 0.65$. Note that only the CLEAN results are blurred here (with the 8.1 GHz CLEAN beam), the \texttt{Comrade.jl} results are shown at their native resolution, which display a significant amount of super-resolution. We show insets to highlight the super-resolved structure near the core. We recover the standard $\alpha>0$ core and $\alpha<0$ jet structure predicted by \citep{BlandfordKonigl1979_core_jet}. However the recovered structure is far more certain with MFS than single frequency imaging.}
    \label{fig:OJsummary}
\end{figure*}

Similar to our validation tests, we performed both single and multifrequency imaging tests to compare the performance of the multifrequency method. Our multifrequency models and priors were the standard ones defined in \autoref{sec:model} and \autoref{sec:priors}. The image FOVs for the OJ287 tests were $20 \times 20 \, \text{mas}$ with image dimensions of $256 \times 256 \,\text{pixels}$. For the multifrequency imaging, the reference frequency was chosen to be 8.1 GHz, and we fixed the total flux of the reference image to be 3.5 Jy. For the single frequency imaging test, we fixed the total flux of the 8.1 and 12.1 GHz images to 3.5 and 5 Jy, respectively. The imaging results are shown in \autoref{fig:OJsummary} in comparison with the published CLEAN results \citep{Hovatta2014_MOJAVE_spectral_index,Lister2018MOJAVE}.

\begin{figure*}
    \centering
    \includegraphics[width=1.0\linewidth]{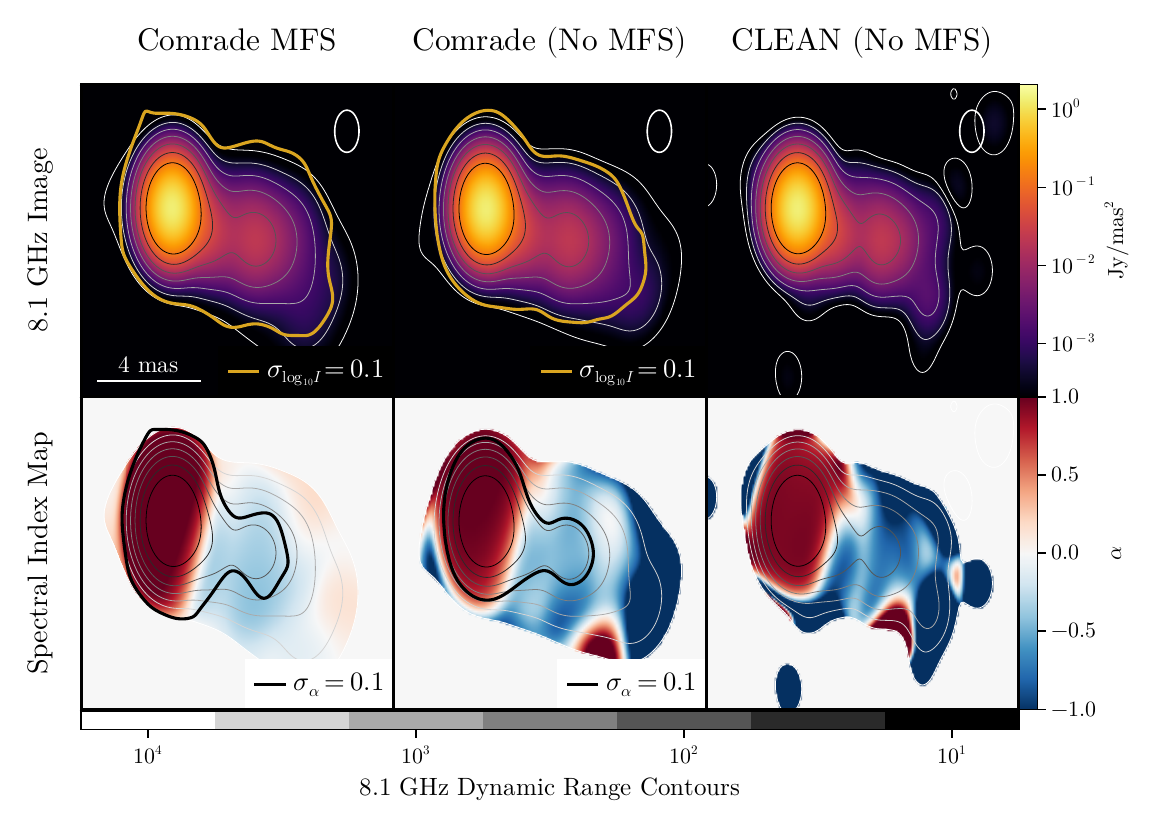}
    \caption{Same as \autoref{fig:OJsummary}, but now all reconstructions are blurred with the 8.1 GHz CLEAN beam. All spectral index maps are derived after convolving the multifrequency 8.1 and 12.1 GHz image samples with the 8.1 GHz CLEAN beam. We demonstrate consistency with the CLEAN total intensity results (see \autoref{fig:OJsuperresolution}). }
    \label{fig:OJsummaryblur}
\end{figure*}

The \texttt{Comrade.jl} imaging results are consistent with the \cite{BlandfordKonigl1979_core_jet} picture of a blazar jet: we recover a point-source core with positive spectral index, and an extended jet feature with negative spectral index. We gain a massive improvement in the quality of the recovered spectral index map when utilizing MFS CHIBI imaging due to the spectral parameterization aligning the image structure across across frequencies. Another striking feature of the OJ287 (C)HIBI reconstructions is that they demonstrate significant super-resolution in comparison to the CLEAN images. Structure below the beam is allowed by the GMRF prior, but is heavily suppressed (see \autoref{sec:model}). Therefore, significant super-resolved structure with high certainty must be driven by the data. We also show these same imaging results blurred to the CLEAN beam in \autoref{fig:OJsummaryblur} to demonstrate consistency with CLEAN.

However, the stark improvement in resolution of Comrade in comparison to CLEAN shown in \autoref{fig:OJsummary} is misleading because of the standard blurring procedure within CLEAN---the CLEAN components may be consistent with Comrade when blurred to scales below the beam resolution. The analysis in \autoref{sec:superresolution} shows consistency between the MFS and total intensity results to $\sim1/3$ of the CLEAN beam.

From \autoref{fig:OJsummaryblur}, the OJ287 multifrequency spectral index map displays what appears to be a systemic offset from the single frequency spectral index maps. This is a fundamental degeneracy in the multifrequency imaging problem present in our current parameterization (see \autoref{sec:discussion} for more details). This offset can be corrected through total flux measurements at both frequencies---we do not correct for it here.

\subsection{MOJAVE: PKS 1424+240}

The BL Lac source PKS 1424+240 is a redshift $z = 0.605$ \citep{Paiano_2017} blazar identified as a potential neutrino source by IceCube \citep{IceCube2022_Neutrino_NGC1068}. Through long-term monitoring with MOJAVE, \cite{Kovalev_2025_PKS1424_240_eyeofsauron} recently showed that the jet has an extremely small viewing angle ($<1^\circ$) within the jet cone, which reveals detailed jet structure. 

We imaged the MOJAVE observations of PKS 1424+240 at 15, 23, and 43 GHz taken with the VLBA on June 26, 2021. Here, we probe a dataset with a much wider frequency range and different source geometry than that of our OJ287 test. Like our OJ287 analysis, we performed a single frequency Comrade fit and MFS Comrade fit, and show our results in comparison with CLEAN results \citep{Kovalev_2025_PKS1424_240_eyeofsauron}.

Our image grid spanned $256 \times 256$ pixels with a FOV of $12\times12$ mas, which corresponds to $3.6$ pixels/beam at 43 GHz---the highest resolution observation in this dataset. We used the same image grid dimensions for all the single and multifrequency image reconstructions. We found that fitting for a spectral index map was sufficient to obtain good fit quality, and therefore we did not perform a spectral index+spectral curvature fit. For the multifrequency fit, we chose the reference frequency to be 15 GHz, and we fixed the total flux at 15 GHz to 0.3 Jy.

\begin{figure*}
    \centering
    \includegraphics[width=1.0\linewidth]{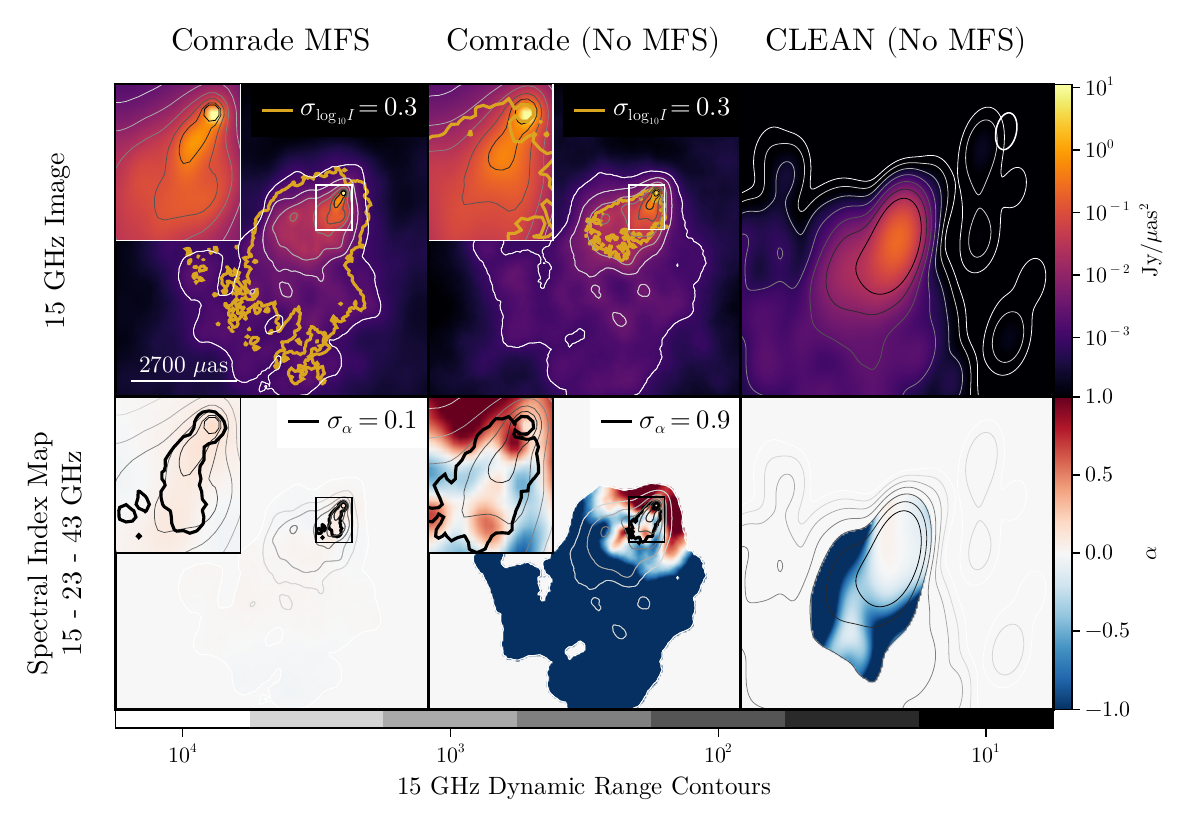}
    \caption{MFS and Single Frequency imaging results of PKS 1424+240 shown in comparison with the correponding published CLEAN \citep{Hovatta2014_MOJAVE_spectral_index,Lister2018MOJAVE} results. We include insets in the Comrade results to highlight the super-resolved core structure. The bold lines represent our credibility contours (gold: $\sigma_{\log I} = 0.3$; black: $\sigma_{\alpha} = 0.9$) and the thin gray-scale lines show logarithmically-scaled total intensity emission for each test. These results demonstrate the floating-point spectral index degeneracy introduced by MFS---without a priori measurements to anchor the total fluxes at 23 and 43 GHz, the MFS spectral index map inherits an unconstrained offset.}
    \label{fig:PKS1424+240}
\end{figure*}

The imaging results are shown in \autoref{fig:PKS1424+240}. As shown in \autoref{fig:OJsummary}, the alignment of structure across frequencies with CHIBI improves the spectral index map. The MFS CHIBI results agree with the relative spectral structure predicted by the \cite{BlandfordKonigl1979_core_jet} model and seen in the prior CLEAN results \citep{Hovatta2014_MOJAVE_spectral_index,Lister2018MOJAVE}: the core has a higher spectral index relative to the downstream jet emission. We do not see a change in sign between the core and jet spectral indices, potentially due to the spectral index offset degeneracy present in MFS imaging (see \autoref{sec:degeneracies} for detailed discussion of this degeneracy).  The considerably improved effective angular resolution of the Comrade.jl images reveal a prominent elongated jet emanating from the core region. The MFS CHIBI results constrain this region of high spectral index to this elongated jet structure emanating from the core, and the low spectral index to the surrounding, diffuse jet emission.

\subsection{Current EHT}\label{sec:EHT}

The current EHT array we consider for this synthetic data test is comprised of the Atacama Large Millimeter Array (ALMA), the Atacama Pathfinder Eperiment (APEX), the Greenland Telescope (GLT), the Institut de Radioastronomie Millimetrique (IRAM), the James Clerk Maxwell Telescope (JCMT), the 12-meter Kitt Peak Telescope (KP), The Large Millimeter Telescope (LMT), the Northern Extended Millimeter Array (NOEMA), the Submillimeter Array (SMA), and the Submillimeter Telescope (SMT). We assume an 8 hour observation during April 1st, with a random weather realization typical of April. These synthetic data were generated by \texttt{ngehtsim} \citep{Pesce2024ngehtsim}.

As test data, we employ theoretical images of M87* at 86, 230 and 345~GHz calculated using general relativistic radiative transfer (GRRT) based on a MAD GRMHD model with a rapidly spinning black hole \citep[the $a_* = 0.9$ simulation from][]{Narayan2022_JetsGRMHD}, which was introduced in \citet{YT2025anisotropy}. In this model, anisotropic nonthermal synchrotron-electrons are injected into the jet, producing a limb-brightened structure consistent with observations of the M87 jet \citep[e.g.,][]{Kim_2018,Walker_2018}, in combination with a helical magnetic-field geometry of the underlying plasma flow. The jet PA was rotated to match the large-scale M87 jet and the total flux density of the images was scaled to match the observed value at 230 GHz \citep{M87PaperIV}. Ten sets of synthetic data were generated at each frequency with an observing track length of 8 hours: these correspond to ten different realizations of April weather. 

\begin{figure*}
    \centering
    \includegraphics[width=1.0\linewidth]{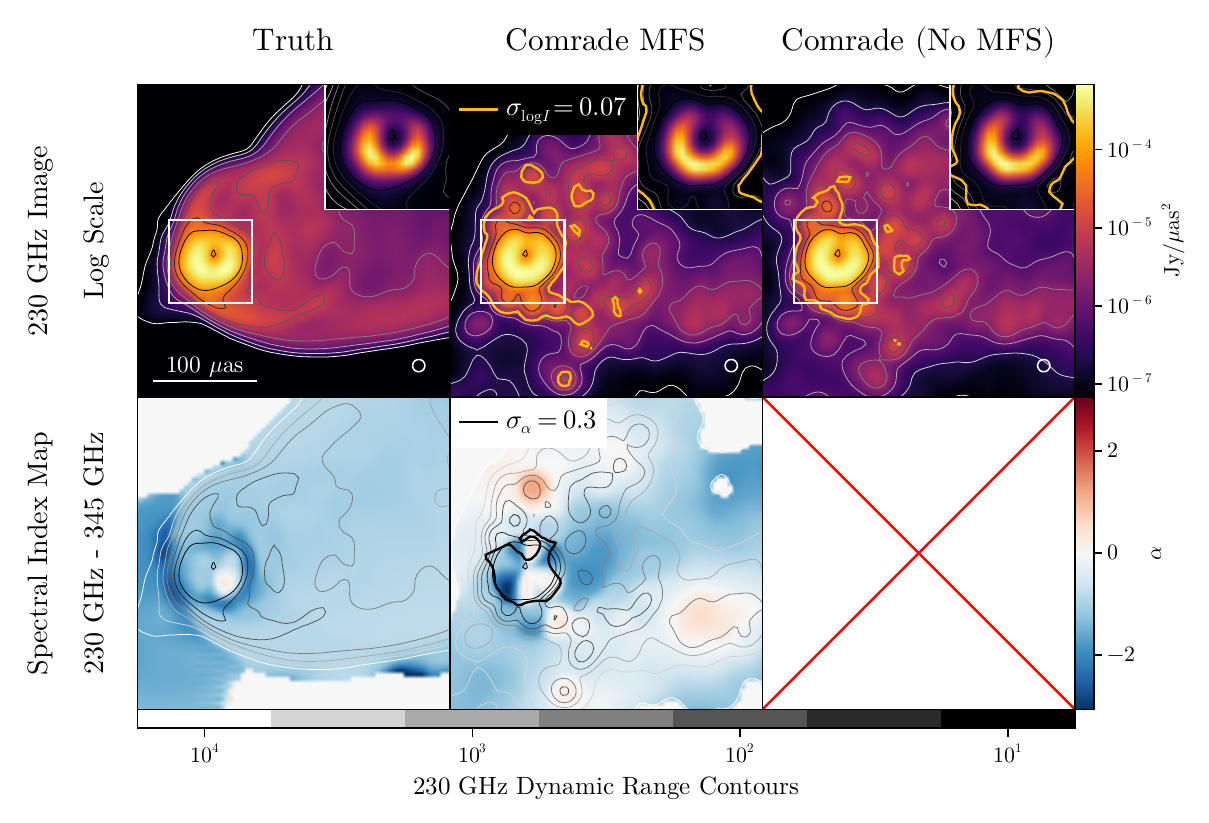}
    \caption{Imaging results of a GRMHD simulation of M87 with the current EHT at 230 and 345 GHz (see \autoref{sec:EHT}). From left to right: the ground truth GRMHD simulation, the CHIBI MFS imaging result, and the Single Frequency HIBI imaging results. The top and bottom rows are the total intensity images (230 GHz) and spectral index maps (230 - 345 GHz). Regions in the MFS spectral index map that correspond to total intensity values less than a factor of $10^{-4}$ dimmer than the maximum brightness have been masked. The insets show the inner ring structure in linear scaling. These images are blurred with a Gaussian kernel with FWHM equal to the \edt{half-beam at 230 GHz ($\sim 12 \, \mu$as)} to increase interpretability of the credibility contours. The solid grayscale contours in each column trace the 230 GHz total intensity emission in log-scale. There is no spectral index map corresponding to single frequency imaging for this dataset---the 345 GHz data are too sparse to image. Producing a spectral index map is only possible with multifrequency synthesis.}
    \label{fig:EHTsummary}
\end{figure*}

\begin{figure*}
    \centering
    \includegraphics[width=1.0\linewidth]{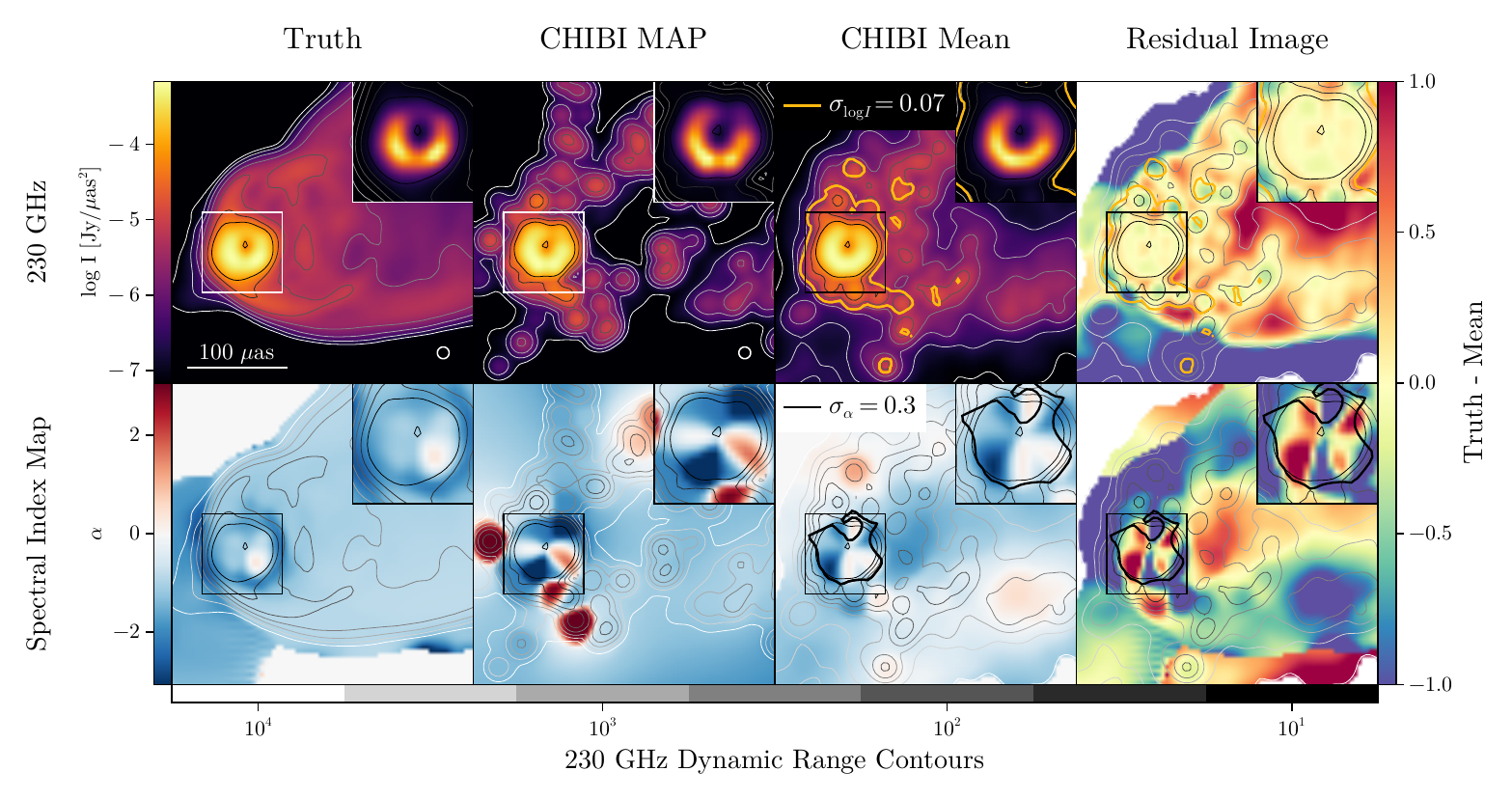}
    \caption{\edt{Metrics validating the EHT multifrequency imaging results with CHIBI. The top and bottom rows show the total intensity and spectral index imaging results, respectively. From left to right: The ground truth image, the MAP image used to initialize the HMC, the mean of the sampled posterior, and the residual image. The insets are shown in linear scaling. The residual images shown are calculated as follows. For total intensity, we calculate the residual between the log-scaled total intensity Truth and CHIBI Mean images: $\log \text{I}_\text{Truth} - \log \text{I}_\text{Mean}$. For spectral index, we calculate the difference images of the spectral index Truth and CHIBI Mean images: $\sigma_\text{Truth} - \sigma_\text{Mean}$. These images are blurred with a Gaussian kernel with FWHM equal to the half-beam at 230 GHz. Although recovery of a spectral index map is enabled by multifrequency synthesis, the undersampling of the data drives uncertainty in the recovered map without additional upgrades.}}
    \label{fig:EHTvalidation}
\end{figure*}

For this test, we did not perform any instrument modeling. We use the image model described in \autoref{sec:model}, with the only variation being a half-normal prior distribution on $\sigma$ being $\mathcal{HN}{(0,0.1)}$ instead of $\mathcal{HN}{(0,0.2)}$. This change placed a more restrictive prior on the image structure and improved sampling performance for this sparse dataset. We performed a multifrequency test with this dataset, and a single frequency test for comparison. Source detections at 345 GHz are more challenging than 230 GHz due to rapid fluctuations in the atmosphere \citep{Pesce2024ngehtsim}. This results in very sparse 345 GHz $(u,v)$ coverage that cannot be imaged via single frequency imaging \autoref{fig:uvcoverage}. As a result, with traditional imaging methods it is impossible to recover a spectral index map for this data set. However, combining the data at both frequencies through MFS allows for the recovery of a 345 GHz and a spectral index map \edt{(see \autoref{fig:EHTsummary} for results and \autoref{fig:EHTvalidation} for validation metrics)}.

\subsection{BHEX}

BHEX is a space-VLBI mission concept which aims to image black holes at unprecedented resolution \citep{Johnson2024BHEX}. The concept is to have a 3.4-meter sub-mm antenna in space observe in tandem with the current EHT. Extending to space would allow BHEX to achieve an angular resolution of $6 \, \mu$as. While BHEX is a proposed mission concept, space VLBI has been successfully demonstrated for decades, and multiple generations of space-VLBI missions have now flown \citep{Levy1986_TDRSS,Hirabayashi1998_HALCA,Kardashev2013Radioastron,Gurvits2020_SVLBI}. Space VLBI represents a unique, but relevant regime in radio astronomy that we explore here with BHEX as an example.

As M87* is one of the BHEX primary targets, we use the same simulated GRMHD snapshot described in \autoref{sec:EHT} as our ground truth image for this test. We assume a BHEX array observing at 79 and 230 GHz, where observations at 79 GHz-only are performed with GBT, IRAM, JCMT, and LMT, observations at 230 GHz-only were made by ALMA, GLT, KP and SMA, and simultaneous dual-band observations at 79+230 GHz are made with the IRAM 30-m, JCMT, and LMT. The stations observing at 79+230 GHz simultaneously are assumed to use frequency phase transfer \citep[FPT; see][and references therein]{Rioja_2020}. This technique leverages the correlated atmospheric phase structure across frequencies to correct for rapid phase fluctuations at higher observing frequencies, and has been demonstrated with VLBI up to 230\,GHz \citep{Issaoun_2025,Zhao_2025}.

We assume a dataset comprised of thirty nights of observation with a three-day cadence, starting on January 1st, 2032. The data are simulated assuming good weather conditions across this observing window. We assume participation from ALMA only during March. For a more detailed instrument and observational cadence description of BHEX, see \citet{Johnson2024BHEX} and \citet{Issaoun2024_BHEX_Operations}.

For our image reconstructions here, we solely model the image structure (no instrument modeling). The image reconstructions had a FOV of $1 \times 1 \, \text{mas}$ and image dimensions of $256 \times 256 \, \text{pixels}$. For the multifrequency fit, we chose 79 GHz as our reference frequency and set the total flux at 79 GHz to be $F_\text{tot, ref} = 1.2 \,$Jy. The total flux densities of the single frequency images at 79 and 230 GHz were set to $1.2 \, \text{Jy}$ and $1.7 \, \text{Jy}$, respectively. Note that these are the ground truth values. 

\begin{figure*}
    \centering
    \includegraphics[width=1.0\linewidth]{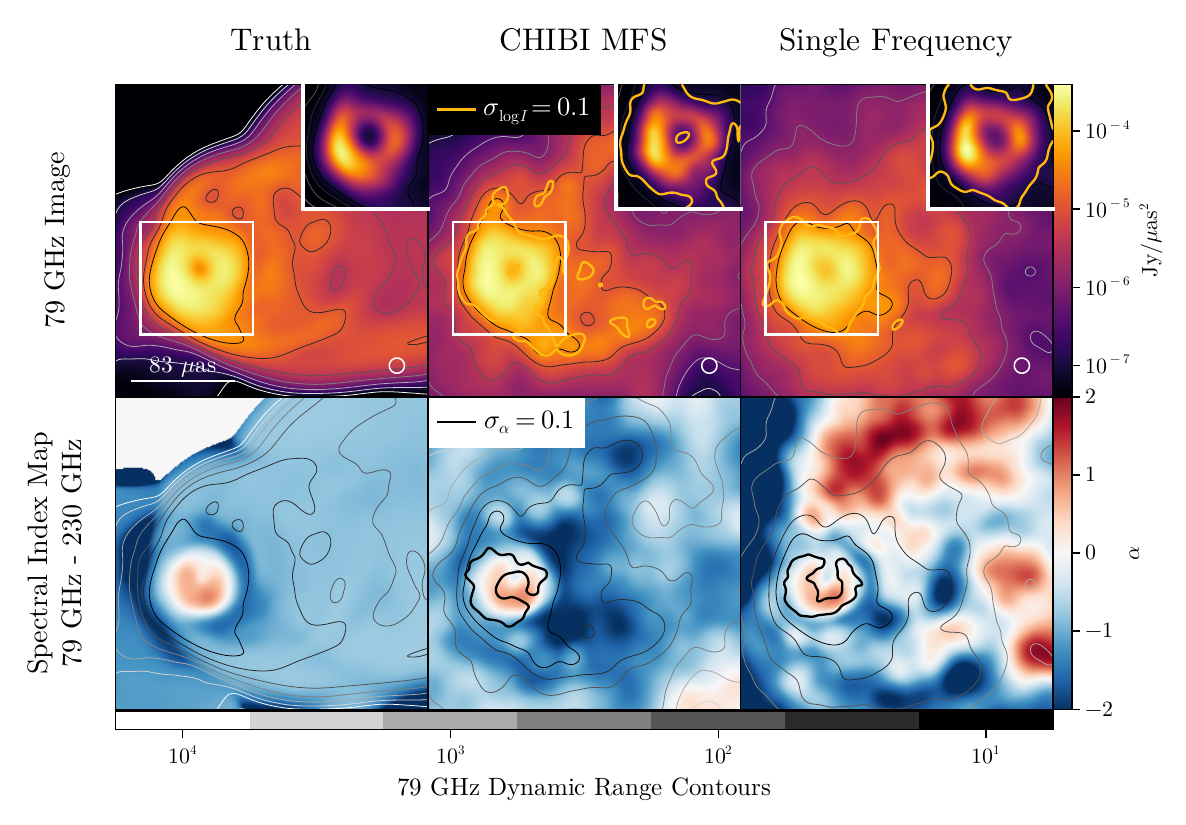}
    \caption{Identical to \autoref{fig:EHTsummary}, but now displaying the imaging results from BHEX + ground array observations of M87 at 79 and 230 GHz. The credibility contours for the total intensity images shown here correspond to $\alpha_{\log I} = 0.1$ (gold), and $\alpha_{\alpha} = 0.3$ (black). \edt{The images are blurred with a circular Gaussian with FWHM equal to half the 79 GHz beam} to increase interpretability of the credibility contours. The high-frequency space baselines contributed from BHEX at 230 GHz contributes a better grasp on the sharp features of the central ring and jet launching region, while the 79 GHz coverage + ground array constrains the jet spectral index. }
    \label{fig:bhexsummary}
\end{figure*}

\begin{figure*}
    \centering
    \includegraphics[width=1.0\linewidth]{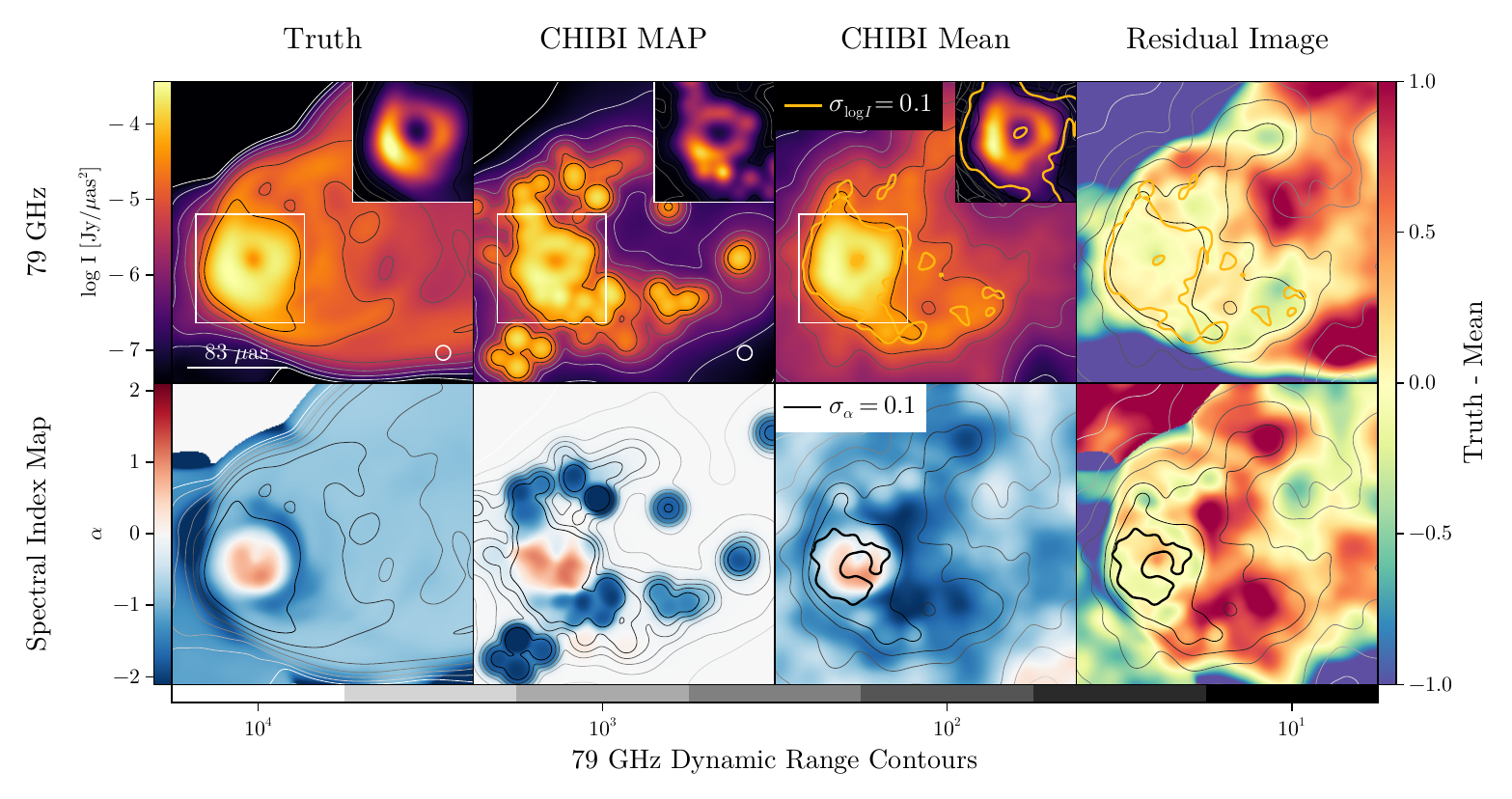}
    \caption{\edt{Identical to \autoref{fig:EHTvalidation}, but now for the BHEX synthetic data test. The images are blurred with a circular Gaussian with FWHM equal to half the 79 GHz beam. Multifrequency synthesis of the combined ground+space coverage improves the recovery of spectral index structure in the ring and jet, however a denser ground array is necessary to accurately recover extended jet features.}}
    \label{fig:BHEXvalidation}
\end{figure*}

We show our imaging results \edt{in \autoref{fig:bhexsummary} and the corresponding validation metrics in \autoref{fig:BHEXvalidation}}. In this test, multifrequency synthesis primarily improves the jet structure. Large and fine scale jet-structure at 79 GHz are both recovered more accurately with multifrequency imaging due to the addition of the 230 GHz ground array and space baselines, respectively. Similarly, while both single and multifrequency imaging both can recover the central ring's spectral index, the jet spectral structure is radically improved with the addition of MFS, emphasizing the potential of multifrequency imaging over a wide frequency range with space VLBI, especially in AGN jet imaging. 

\subsection{ngEHT\label{sec:ngEHT}}

The ngEHT is an array design effort to upgrade the EHT telescopes with simultaneous multi-frequency wide-bandwidth systems and to motivate additional telescopes at key sites that enable broader (sub)millimeter black hole science with an enhanced and denser array \citep{Johnson2023_ngEHT,Doeleman2023_ngEHT}. 

We assume the ngEHT phase 2 array configuration and frequency capabilities from Table A2 of the ngEHT reference array paper \cite{Doeleman2023_ngEHT}, minus BOL, KILI, KVN stations, LLA, SGO, SPT, and SPX for a total of 17 sites. Here we assume that ALMA and SMA observe only at 230 GHz, all other stations simultaneously observe at 86 and 230 GHz \citep{Issaoun_2023}. The data are generated with weather conditions typical of April observations. We used the 86 and 230 ray-traced images of the M87 GRMHD simulation described in \autoref{sec:EHT} as the ground truth images for this test. The image raster grid spans $256 \times 256$ pixels and has an image FOV of $1.8 \times 1.8 \,$mas. We chose 86 GHz as our reference frequency for the multifrequency image reconstruction, and fixed the 86 GHz total flux density to 1.7 Jy. The total flux densities at the 86 and 230 GHz single frequency images are 1.7 and 1.2 Jy, respectively.

Here we note one change in the image model that we made for this test. We demonstrate an alternative parameterization for the spectral index map $\alpha(x,y)$ that allows the spectral-index offset degeneracy to be sampled more directly (see \autoref{sec:discussion} for more details). For this test, we model the spectral index $\alpha(x,y)$ as
\begin{equation}
    \alpha(x,y) = \alpha_0 + \sigma_\alpha \alpha'(x,y).
    \label{eq:fancyspectralindex}
\end{equation}
This simplifies to the standard spectral index model described in \autoref{sec:model} when $\alpha_0 = 0, \sigma_\alpha = 1$.

In this alternative parameterization, $\alpha_0$ represents the zero-point value of the spectral index map, and $\sigma_\alpha$ is a dimensionless scaling factor to modify the spectral index map fluctuations: $\alpha'(x,y)$. We maintain the same GMRF prior on $\alpha'(x,y)$ described in \autoref{sec:priors}:
\begin{equation}
    \bm{\alpha'} \sim {\rm GMRF}[0, \bm{Q}_1(\sigma=1)]
\end{equation}
We set the prior on $\alpha_0$ to be a normal distribution with mean 0 and standard deviation 0.1:
\begin{equation}
    \bm{\alpha_0} \sim \mathcal{N}(0,0.1).
\end{equation}
The prior on $\sigma_\alpha$ is a half-normal distribution with mean 0, standard deviation 0.01, and lower bound of 0:
\begin{equation}
    \bm{\sigma_\alpha} \sim \mathcal{HN}(0, 0.01).
\end{equation}

\begin{figure*}
    \centering
    \includegraphics[width=1.0\linewidth]{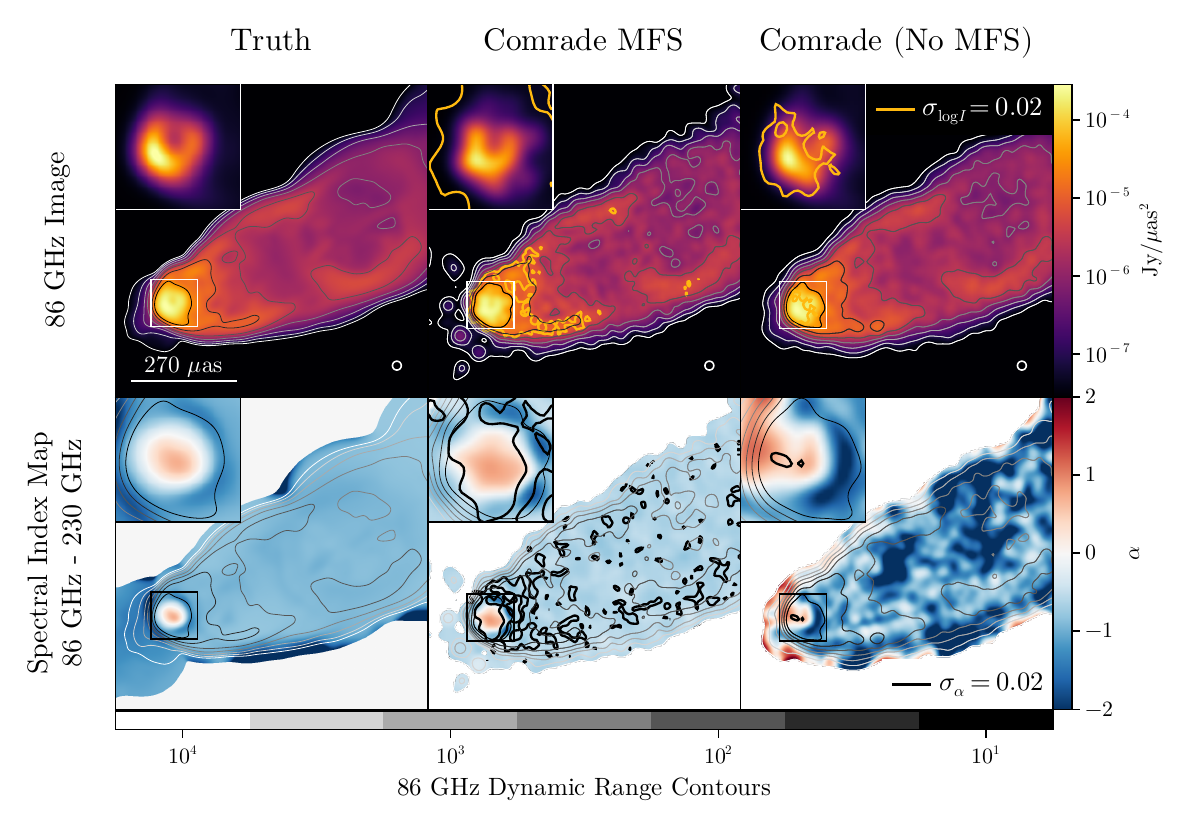}
    \caption{The same as Figures \autoref{sec:EHT} and \autoref{fig:bhexsummary}, now for simulated ngEHT observations of M87 at 86 and 230 GHz. The images are blurred \edt{with one-third of the ngEHT 86 GHz beam} for interpretability of the credibility contours. The bolded credibility contours denote $\sigma_{\log I } = 0.02$ (gold), and $\sigma_{\alpha} = 0.02$ (black). Here, the spectral index insets retain the same scaling as the full spectral index maps. Here, the primary improvement with MFS is shown in the spectral index map---correlating the structure between the two frequencies through multifrequency modeling greatly improves recovery of the spectral index at all scales.}
    \label{fig:ngEHTsummary}
\end{figure*}

\begin{figure*}
    \centering
    \includegraphics[width=1.0\linewidth]{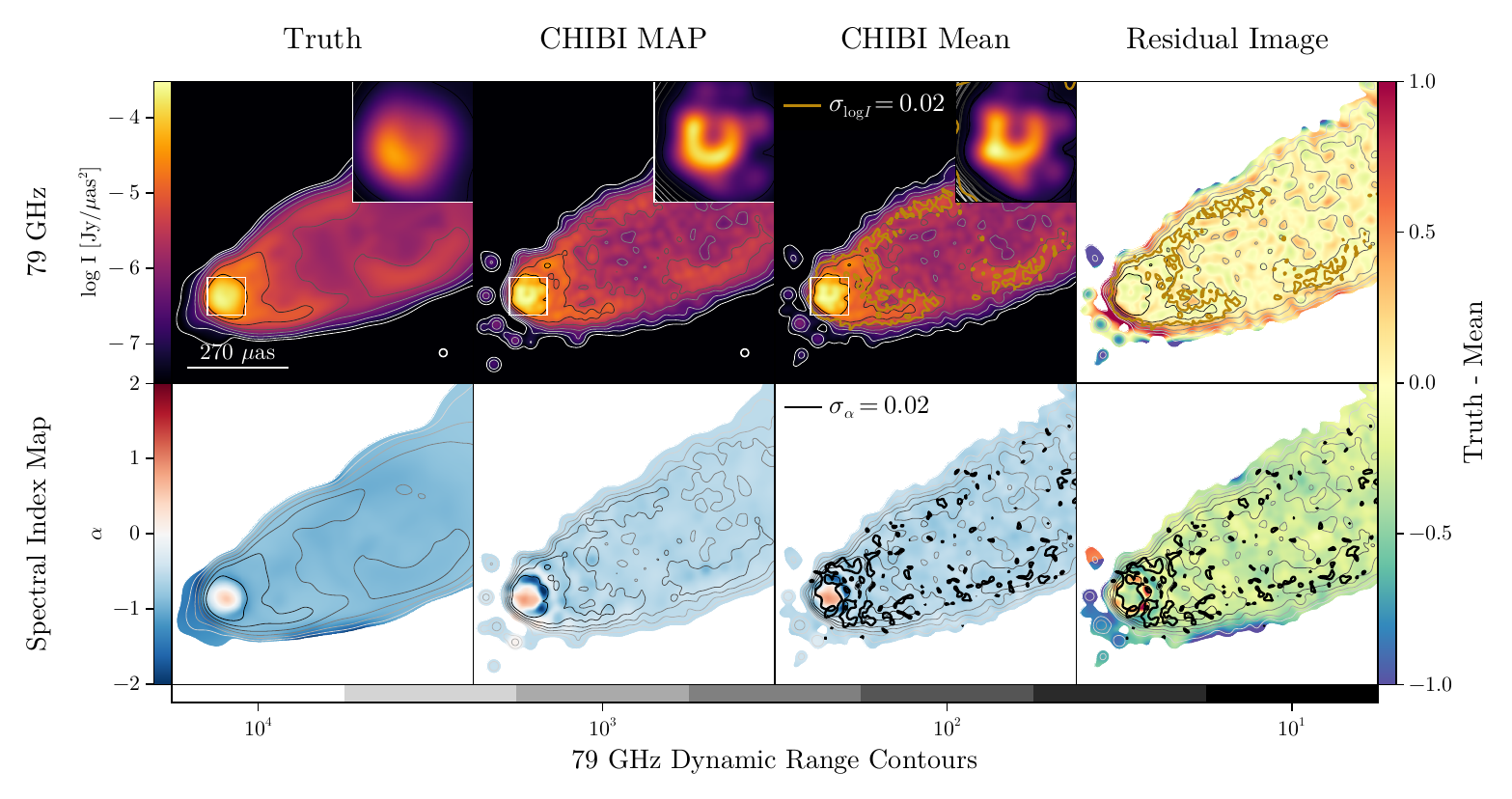}
    \caption{\edt{Identical to \autoref{fig:EHTvalidation} and \autoref{fig:BHEXvalidation}, but now for the ngEHT synthetic data test. All images are blurred with one-third of the ngEHT 86\,GHz beam. Multifrequency synthesis with the dense ground array enables high-fidelity recovery of extended jet structure in both total intensity and spectral index. Here we see the effect of the spectral index degeneracy described in \autoref{sec:degeneracies}: a uniform offset in the residual map.}}
    \label{fig:ngEHTvalidation}
\end{figure*}

The imaging results \edt{and corresponding validation metrics are shown in \autoref{fig:ngEHTsummary} and \autoref{fig:ngEHTvalidation}}. This test emphasizes the value of MFS in terms of image alignment. The ngEHT array is already dense at both frequencies, and therefore the improvement in the total intensity image at 86 GHz with MFS is subtle. However, the improvement in spectral index is striking---this improvement arises from MFS aligning the spectral index structure at both frequencies. As discussed in \autoref{sec:discussion}, the spectral index GMRF prior prefers aligned structure between frequencies---this produces clean, interpretable spectral index maps. The alignment is further technically supported by the simultaneous multi-frequency design of the instrument itself. 

\section{Discussion} \label{sec:discussion}

\subsection{Image Degeneracies}\label{sec:degeneracies}

There are a variety of fundamental degeneracies that arise during the VLBI imaging problem, some of which are unique to multifrequency synthesis. 

There is a total flux degeneracy in VLBI imaging. When modeling instrumental gains (analogous to self-calibration) or when fitting closure amplitudes as our data product, the total flux of the image is degenerate. In the case of modeling instrumental gains, the image total flux can be changed via the gain amplitudes $A_p$ (see \autoref{eq:gains}) and still retain consistency with data. Log-gain priors centered around 0 weakly breaks the total flux degeneracy, but the existence of any prior mass on the gains leaves some degeneracy still open. Likewise, by definition, closure amplitudes are invariant to a change in total flux, so the degeneracy remains even if we circumvent gain amplitude modeling by using closure products as our data product.

Another degeneracy is an unconstrained image centroid. This emerges in two different situations: when modeling instrumental gain phases and when fitting closure phases. In both cases only the \textit{relative} phase structure is preserved---an absolute phase shift can be absorbed into the instrumental gain phases, and closure phases are inherently invariant to a global phase shift. An absolute phase difference will shift the entire image, thus, the image centroid location is unconstrained.

In the context of multifrequency imaging, each individual frequency band suffers from the aforementioned degeneracies, resulting in a lack of knowledge about the relative flux densities and relative centroids of each band with respect to every other band. When modeling gains or fitting closure products, now the total flux at \textit{each frequency} is unconstrained. This produces an unknown overall offset in the spectral maps---only the relative spectral structure is constrained. Unconstrained \textit{relative} image centroids similarly naturally follow from each image having unconstrained centroid locations---the relative centroids at each frequency are unconstrained as well.

These issues compound when fitting for additional frequencies of data---each additional frequency introduces a potential centroid offset---and when fitting for higher order spectral structure. Each order (spectral index, spectral curvature, etc...) introduces an unknown spectral offset.

\subsection{Addressing the Degeneracies \label{sec:solutions_multimodal}}

These various degeneracies complicate posterior structure by introducing correlations and multi-modality (multiple, distinct image solutions) which make posterior sampling intractable. We address this by parameterizing the image model such that a priori source-specific information can break these degeneracies. 

For the single frequency imaging model, we explicitly parametrize the total image flux density. Therefore, we can use a priori flux measurements to break the degeneracy (see $F_{tot}$ in \autoref{eq:totalflux}). Uncertainty on the total flux measurement can be folded in via a prior on $F_{tot}$. For the multifrequency imaging model, only the reference image has a parameterized total flux density. The total flux density at other frequencies are free parameters. As a result the spectral offset described in \autoref{sec:degeneracies} remains---only the relative structure is constrained. 

To address this challenge, our standard treatment is to set the mean structure of the spectral GMRF prior to be a flat value across the entire image. Our prior on this mean value is a normal distribution $\mathcal{N}(0,1)$. This places the bulk of the probability mass within the range of spectral map values expected for synchrotron radiation. This is the spectral index model we used in our Gaussian validation test, the MOJAVE data, and on synthetic data generated for the current EHT and BHEX tests. 

In our ngEHT test in \autoref{fig:ngEHTsummary}, we parameterize the spectral offset more explicitly. We model the mean GMRF prior for the spectral structure as \autoref{eq:fancyspectralindex}: $\alpha(x,y) = \alpha_0 + \sigma_\alpha \alpha'(x,y).$
Here, ($\alpha_0$) is an overall offset in the spectral map that we sample. Explicitly modeling the offset makes it simpler for the HMC sampler to explore this offset value. However, the degeneracy isn't truly broken because the fluctuations superimposed upon this mean structure also determine the total flux density of the images at other frequencies, but it can be explicitly broken with a priori total flux measurements at each frequency

As mentioned in \autoref{sec:degeneracies}, modeling instrumental gain phases results in an unconstrained image centroid. In \autoref{sec:model}, we presented this issue as a result of the image structure drifting between posterior samples. Our solution is outlined in \autoref{sec:model}: we reduce the size of the Gaussian distribution used for the mean structure of the GMRF prior to concentrate the image flux in a limited region of the image, which restricts image drift. Ultimately, image drift cannot be completely eliminated in our model. Therefore when calculating statistical quantities from the posterior (e.g. mean, standard deviation), we choose to center all the VLBA Blazar observations with a bright, unresolved core with regards to the brightest point in the image (the core), and all other images with respect to their image centroids.

For our multifrequency model, centering the reference image and corresponding spectral maps with regard to the reference's brightest pixel or image centroid does not correct for the \textit{relative} offset between images. There are two ways to address this: in the image model, or in the instrument model. Currently we only address it with the image model, but we outline here how future instrument model development would present a superior solution.

If we impose a particular image geometry (e.g. a ring), an image offset could be explicitly parameterized. However, since we do not assume specific image geometry, we rely on the spectral image GMRF prior to align the images. Misaligned image structures cause large fluctuations in spectral index across the entire image. Therefore, our standard image prior which prefers a flat spectral map around 0 will preferentially align the images. 

In the case where any of these degeneracies cause multi-modal image structure (e.g. an image flip, multiple possible image alignments), as mentioned in \autoref{sec:initialization_data_products}, our NUTS sampler may fail to explore each mode during a simgle sampling run. Therefore, multiple image runs with a variety of distinct initializations is typically necessary to discover the various image modes. \edt{For each of our tests shown here, we performed multiple fits with different independent initializations. High-certainty regions were statistically consistent across the different runs. In low SNR regions of the image reconstructions, features were recovered up to expected MC error. In the cases where we did see discrepancies between initializations, these were associated with the degeneracies discussed in \autoref{sec:degeneracies}.}

\subsection{Future development}
To explicitly break the relative image position degeneracy, we can use novel observational techniques such as frequency phase transfer \citep[FPT; e.g.,][]{Rioja_2015,Rioja_2023} and source frequency phase referencing \citep[SFPR; e.g.,][]{Zhao_2019,Rioja_2020}, which constrain the relative phase structure between multiple frequencies. FPT has recently been successfully demonstrated for high-frequency VLBI and will enable MFS with EHT-like arrays in the future \citep{Issaoun_2025, Zhao_2025}. This technique will automatically align our multifrequency images --- errors in this alignment will naturally arise from phase errors in the data. This requires the future development of a multifrequency instrument model with correlated phase structure across multiple frequencies.

Computationally, there are avenues to improve performance and speed. Parallelization and GPU compatibility will be essential to enable MFS Comrade for larger datasets than VLBI, such as phased arrays. 

Algorithmically, we only tested a single spectral model implementation in this paper---a logarithmic polynomial expansion. For emission mechanisms where this polynomial expansion is purely phenomenological (i.e., not physically well-motivated), the code is modular enough that an interested user can implement alternative spectral models. Future development in this area may include implementing a multi-gaussian spectral line model to perform spectral-line analysis.

\section{Conclusions}

We presented a new method for multifrequency synthesis in a Bayesian inference framework, implemented in \texttt{Comrade.jl}. The model parametrization is rooted in the spectral behavior of synchrotron radiation, the primary radio emission mechanism in supermassive black holes. We validated the performance of our Bayesian multifrequency synthesis implementation and found that it is able to recover the source multifrequency structure more accurately and precisely than single frequency imaging. In addition, we found that the multifrequency structure derived from MFS is clearly data-driven.

We tested MFS with a variety of VLBI arrays representing various approaches and challenges to multifrequency imaging. We applied our MFS technique to real data of AGN blazars (OJ287, 1424+240) and then to synthetic observations of M87*'s near-horizon structure with the current EHT and future arrays, such as the Black Hole Explorer space-ground interferometer and the next-generation EHT \citep{Johnson2024BHEX,Doeleman2023_ngEHT}. 

The structure recovered by Comrade MFS with the VLBA observations is consistent with the published CLEAN imaging results \citep{Hovatta2014_MOJAVE_spectral_index,Lister2018MOJAVE}, with clear improvements in image resolution and recovery of structure in the spectral index maps compared to CLEAN. For the current EHT, MFS is required to create a 345 GHz image, and, consequently, a 230-345 GHz spectral index map, due to the sparsity of the coverage at that frequency. The improved coverage enabled by BHEX allows imaging at both simulated frequencies (79 GHz and 230 GHz). The ($u,v$) coverage added by the space baselines at 230 GHz greatly improves the image resolution at 79 GHz, and the 79 GHz baselines help constrain the jet emission and spectral index. For the ngEHT result, the greatest improvement is shown in the spectral index map. We attribute this to the dense $(u,v)$ coverage, but also the image alignment of structure between frequencies enforced by MFS. In all cases, MFS improved the total intensity images and the resulting spectral index maps. This is due to the improvement in $(u,v)$ coverage from combining multiple datasets sampling different spatial frequencies, and also due to the alignment of image structures across multiple frequencies. This method will greatly improve spectral index measurements and spatially-resolved spectral index maps of black hole jets with VLBI. 

The current implementation of our Bayesian MFS framework can be applied to a variety of scenarios. As of 2018, the EHT observing setup has four sub-bands at 213, 215, 227, and 229~GHz, with existing observations \citep{EHT2018M87PI,EHT2021M87}. No modification to the model is necessary to fit sub-band observations, which are also truly simultaneous, making the most recent EHT observations ideal datasets for MFS. Future work will also include multifrequency analysis of non-AGN continuum emission, such as protoplanetary disks \citep[e.g.,][]{HLTau_2015}. While we focused on VLBI arrays in this analysis, this technique can also be applied to data from phased interferometers, such as ALMA or SMA. The primary barrier here is the computational cost due to much larger data volumes than the VLBI arrays we tested in this paper. 

Further development of our Bayesian MFS framework involves extending to polarized multifrequency synthesis. In addition to improving the polarized image structure at each frequency, this would enable the recovery of Faraday rotation measure maps as a result of the imaging process \citep[e.g.,][]{Brentjens_2005,Andrecut_2012,Bell_2012}. Rotation measure maps are particularly important to infer magnetic field structure and plasma properties of AGN jets and accretion flows \citep[e.g.,][]{Gabuzda_2004,Hovatta_2012,Marrone_2007,Bower_2018}. The Bayesian framework would enable uncertainty quantification on these rotation measure maps as well. 

Implementing additional spectral models into the multifrequency model would be another natural extension to this method. Currently, the polynomial expansion model we have implemented can only model continuum radio emission. With the addition of a multi-component Gaussian spectral model, we can simultaneously model continuum and spectral-line emission. Our framework would then be useful for a variety of other radio targets exhibiting, e.g., maser line emission or protoplanetary disk molecular line emission.

Future work will also incorporate multifrequency instrument modeling. The current instrument model is appropriate for observations where the instrumental corruptions and atmospheric corruptions are uncorrelated between the different frequencies. This works well for quasi-simultaneous observations---non-simultaneous multifrequency observations where we assume the source is stationary and the atmosphere variability is largely uncorrelated. However, truly simultaneous multifrequency observations with frequency phase transfer and its variants introduce correlations in the gain structure across frequencies \citep[e.g.,][]{Asaki_1996,Asaki_1998,Dodson_2017,Rioja_2014,Rioja_2015,Zhao_2018}. Modeling this frequency-dependent gain structure explicitly would break a variety of image degeneracies in the multifrequency problem, such as the relative image position degeneracy (for FPT), absolute image position degeneracy (for source frequency phase referencing), and the overall spectral index offset. It also opens the possibility for Bayesian multifrequency astrometry, of which one application is measuring AGN core-shifts \citep{Rioja_2020}.

\newpage
\newpage
\newpage

\section*{Acknowledgements}

We acknowledge financial support from the National Science Foundation (AST-2307887, AST-1935980, AST-2034306). This publication is funded in part by the Gordon and Betty Moore Foundation, Grants GBMF-12987, GBMF-5278, GBMF-10423. This work was supported by the Black Hole Initiative, which is funded by grants from the John Templeton Foundation (Grant \#62286) and the Gordon and Betty Moore Foundation
(Grant GBMF-8273)---although the opinions expressed in this work are those of the author(s) and do not necessarily reflect the views of these Foundations. This material is based upon work supported by the National Science Foundation Graduate Research Fellowship under Grant No. DGE 2140743.

\vspace{5mm}

\edt{\software{Comrade.jl \citep{tiede2022}, 
          ehtim \citep{chael2023ehtim},
          Makie.jl \citep{DanischKrumbiegel2021}}}

\appendix
\section{Super-resolution Analysis: OJ287}\label{sec:superresolution}
The CHIBI and HIBI imaging results of the OJ287 MOJAVE data shows significant super-resolution. To determine this, we perform a super-resolution consistency test between the three images: CLEAN, Single Frequency Comrade, and MFS Comrade. Following previous super-resolution studies to compare different imaging algorithms \citep[e.g.,][]{M87PaperIV}, we use the normalized cross-correlation (NXCORR) as our comparison metric:
\begin{equation}
    \rho_{NX}(X,Y)=\frac{1}{N}\sum_i \frac{(X_i-\langle X\rangle)(Y_i-\langle Y\rangle)}{\sigma_X\,\sigma_Y}.
    \label{eq:NXCORR}
\end{equation}
Here, $X_i$ and $Y_i$ are the pixel values of the two compared images, $\sigma_X$, $\sigma_Y$ are the standard deviation values of the pixel values from each image, and $N$ is the total number of pixels in the image.

\begin{figure*}
    \centering
    \includegraphics[width=1.0\linewidth]{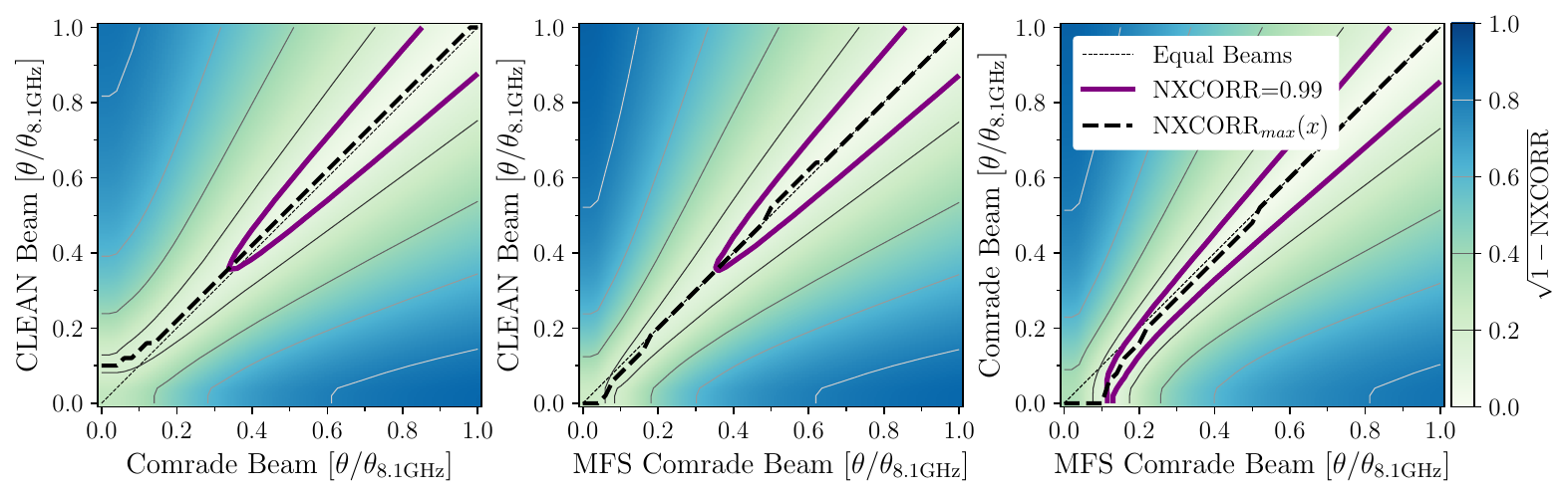}
    \caption{A NXCORR analysis which compares the similarity of the super-resolved structure from the total-intensity CLEAN \citep{Lister2018MOJAVE}, single frequency Comrade, and multifrequency Comrade image reconstructions of OJ287 at 8.1 GHz. The axes represent the FWHM of the Gaussian blurring kernel applied to each image---they are normalized to the 8.1 GHz array resolution ($\theta_\text{8.1 GHz}$). We use a threshold value of NXCORR=0.99 (purple contour) to determine that two images are identical---the CLEAN and Comrade reconstructions of OJ287 are consistent to about 1/3 of the CLEAN beam.}
    \label{fig:OJsuperresolution}
\end{figure*}

Our procedure included blurring each of these image reconstructions with circular Gaussian beams whose FWHMs ranged from 0 (no blur) to the array resolution, and computing the NXCORR between each pair of images for each combination of blurring kernel size between pairs of images. Note that the CLEAN beam is highly elliptic because of the asymmetric $(u,v)$ coverage of this observation. For the Comrade and CLEAN images we blurred the raw images (shown in \autoref{fig:OJsummary}) and the CLEAN components, respectively. We choose the threshold value of NXCORR=0.99 to determine that the compared images are identical. The results are shown in \autoref{fig:OJsuperresolution}, which demonstrates that the CLEAN and Comrade images are consistent to $\sim 1/3$ of the CLEAN beam.

\section{MOJAVE Images at all Frequencies}

\begin{figure*}
    \centering
    \includegraphics[width=\linewidth]{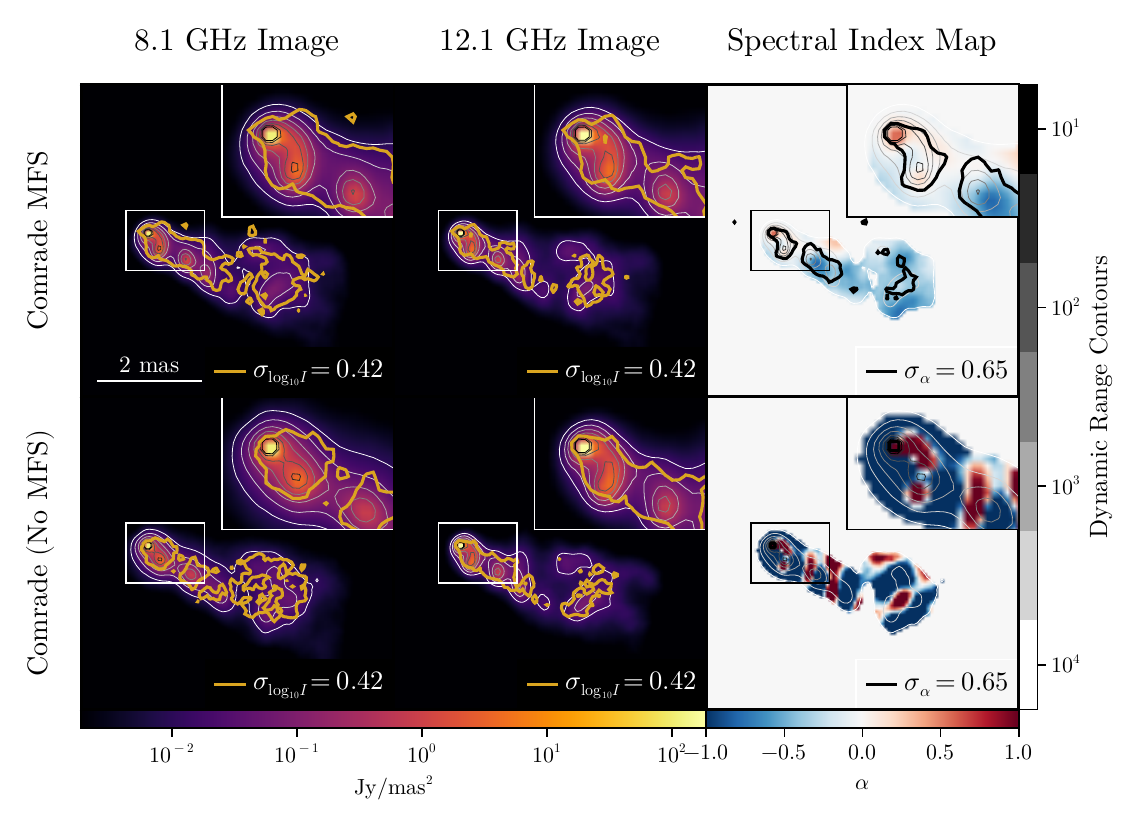}
    \caption{OJ287 VLBA MOJAVE image reconstructions at all frequencies (8.1, 12 GHz) and the resulting spectral index maps. See \autoref{fig:OJsummary} for full description of all figure elements.}
    \label{fig:OJappendix}
\end{figure*}

\begin{figure*}
    \centering
    \includegraphics[width=\linewidth]{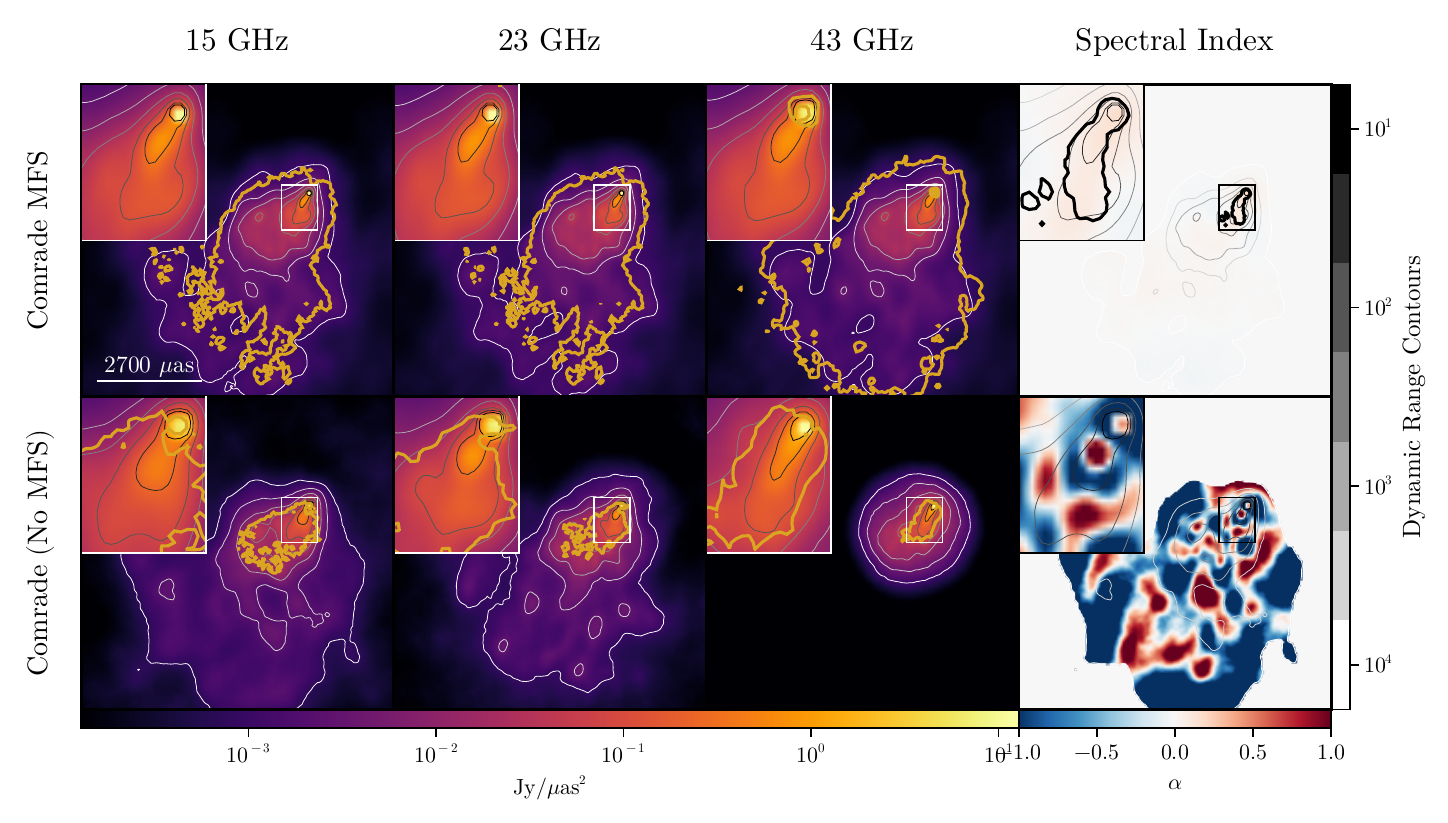}
    \caption{PKS 1424+240 VLBA MOJAVE image reconstructions at all frequencies (15, 23, 43 GHz) and the resulting spectral index maps. See \autoref{fig:PKS1424+240} for full description of all figure elements. Note that at the highest frequency of 43 GHz, the extended jet emission is only recoverable via multifrequency imaging.}
    \label{fig:1424+240appendix}
\end{figure*}

In \autoref{fig:OJappendix} and \autoref{fig:1424+240appendix} we show the imaging results at all frequencies of data from MFS CHIBI and single frequency HIBI imaging of OJ287 and PKS 1424+240. 

\section{Goodness of Fit for Image Reconstructions}
\begin{figure*}
    \centering
    \includegraphics[width=1\linewidth]{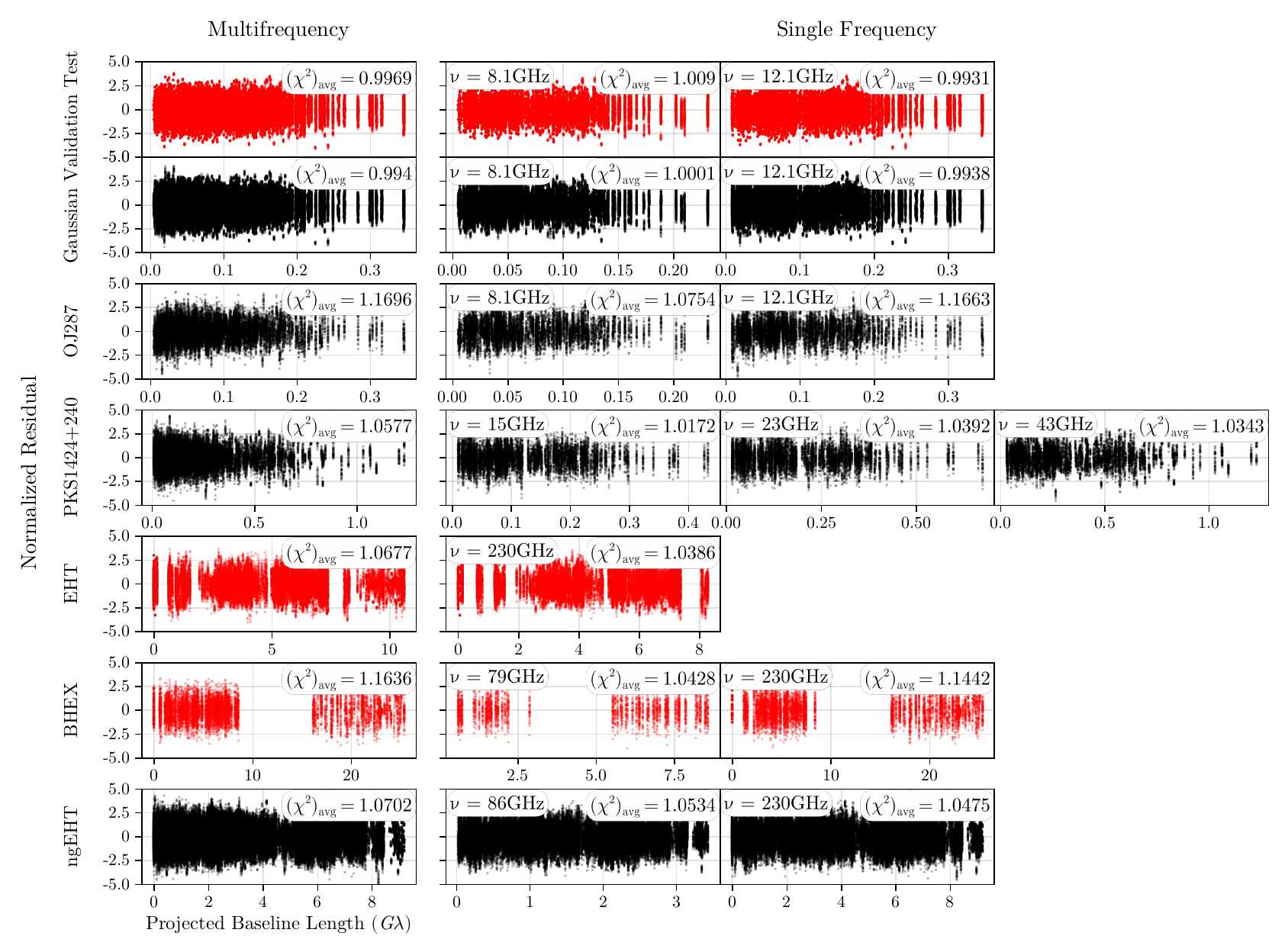}
    \caption{The normalized Residuals, aggregated across 20 random draws from the posterior, for all imaging tests performed in this paper. The averaged $\chi^2$, $\chi^2_{avg}$ across each of those 20 samples are over-plotted for each test. The tests which perform joint image and instrument modeling are colored in black, tests which solely perform image modeling (no instrument modeling) are colored in red. We plot both the real and imaginary components of the normalized complex visibility residual on each plot.}
    \label{fig:chi2}
\end{figure*}

We show goodness of fit statistics for each imaging test shown in this paper in \autoref{fig:chi2}. We calculate the normalized residual as the difference between the measured and modeled complex visibility, normalized by the measurement noise. The normalized residual is calculated separately for the real and imaginary components of the complex visibilities.

\bibliography{references}{}
\bibliographystyle{aasjournal}

\end{document}